\documentclass[aps,prd,preprintnumbers,nofootinbib,floatfix,floats,superscriptaddress]{revtex4}%{,twocolumn}
\usepackage{bm}
\usepackage[utf8]{inputenc} 
\usepackage{latexsym}
\usepackage{dcolumn}
\usepackage{amsmath,amsfonts,amssymb,mathtools}
\usepackage{empheq}
\usepackage{graphicx,epsfig}
\usepackage{float}
\usepackage{color}
\usepackage[active]{srcltx}%DO NOT DELETE /
\usepackage[caption=false]{subfig}
\usepackage{slashed}
\usepackage{hyperref}
\usepackage{comment}
\usepackage{tikz}
\usetikzlibrary{shapes,snakes}
\usepackage{multirow}
\usepackage[normalem]{ulem}%for strikethrough
\usepackage[mathscr]{euscript}
\usepackage{mathrsfs}
\usepackage{enumitem}

\def\l{\left}
\def\r{\right}
\def\DM{\mathrm{d}}

\newcommand{\gae}{\lower 3pt \hbox{$\,\, \buildrel {\scriptstyle >}\over {\scriptstyle
\sim}\,\,$}}
\newcommand{\lae}{\lower 2pt \hbox{$\, \buildrel {\scriptstyle <}\over {\scriptstyle
\sim}\,$}}

\DeclarePairedDelimiter\bra{\langle}{\rvert}
\DeclarePairedDelimiter\ket{\lvert}{\rangle}
\DeclarePairedDelimiterX\braket[2]{\langle}{\rangle}{#1\,\delimsize\vert\,\mathopen{}#2}
\begin{document}

%\begin{comment}

\title{Entanglement between accelerated probes in a de Sitter spacetime}

\author{Mayank}
\email{mayank@physics.iitm.ac.in}
\affiliation{Centre for Strings, Gravitation and Cosmology, Department of Physics, Indian Institute of Technology Madras, Chennai 600 036, India}
\author{K. Hari}
\email{hari.k@iitb.ac.in}
\affiliation{Centre for Strings, Gravitation and Cosmology, Department of Physics, Indian Institute of Technology Madras, Chennai 600 036, India}
\affiliation{Department of Physics, Indian Institute of Technology Bombay, Mumbai 400076, India}
\author{Subhajit Barman}
\email{subhajit.barman@physics.iitm.ac.in}
\affiliation{Centre for Strings, Gravitation and Cosmology, Department of Physics, Indian Institute of Technology Madras, Chennai 600 036, India}
\author{Dawood Kothawala}
\email{dawood@iitm.ac.in}
\affiliation{Centre for Strings, Gravitation and Cosmology, Department of Physics, Indian Institute of Technology Madras, Chennai 600 036, India}

\date{\today}
\begin{abstract}

\noindent We initiate an investigation into features of vacuum entanglement as probed by accelerated quantum probes in curved spacetime. Focussing specifically on de Sitter (dS) spacetime with curvature $\Lambda$, we obtain several exact results corresponding to different kinematical set-up of the probes. The interaction with the quantum field creates a non-local correlation between initially uncorrelated probes accelerating in different directions. It is well known that a single quantum probe in dS spacetime with uniform acceleration $a$ responds exactly as a quantum probe in Minkowski spacetime with ``effective" acceleration $q \equiv\sqrt{a^2+\Lambda}$. However, no such mapping generically exists for the entanglement between probes. Our results suggest that entanglement exhibits independent variations with changes in acceleration and curvature depending on different configurations of detector motion.
\end{abstract}

\maketitle

\tableofcontents

%%%%%%%%%%%%%%%%%%%%%%%%%%%%%%%%%%%%%%%%%%%%%%%%%%%%%%%%%%%%%%%%%%
\section{Introduction} \label{sec:intro} 
%%%%%%%%%%%%%%%%%%%%%%%%%%%%%%%%%%%%%%%%%%%%%%%%%%%%%%%%%%%%%%%%%%

Accelerated frames have always played a crucial role in understanding the physical processes happening in a curved background. For instance, phenomena such as the gravitational time dilation and the Hawking effect are better understood in their analogies from accelerated frames in Minkowski spacetime. The Unruh effect is perhaps one of the most well-known physical effects that arises when one considers uniformly accelerated quantum probes coupled to a quantum field. Such probes sample quantum correlations of a quantum field, and respond with a typical black body response at the so-called Unruh temperature which is proportional to the acceleration. However, the effect is miniscule at accessible accelerations and, hence, has yet to be verified directly through observations. Efforts to do so have led to several investigations through which such an effect can be observed at lower accelerations by carefully designing the set-up so as to leverage the modified behaviour of field modes \cite{Stargen:2021vtg}. 

However, if the aim is to just check the validity of the application of conventional QFT to accelerated systems, there are other options that can be explored. If one shifts focus from a single detector to two detectors, both quantum mechanical, there is the possibility that quantum correlations of the field can induce entanglement between the detectors, which can then be presumably measured. This was first pointed out by Reznik \cite{Reznik:2002fz, Reznik:2003mnx}, and since then explored extensively under the name of {\it entanglement harvesting} \cite{Martin-Martinez:2015psa, Tjoa:2021roz}. In this protocol, the entanglement is induced between two initially uncorrelated quantum probes, through their interaction with the background field, which are in different trajectories and are usually considered to be causally disconnected. This entanglement between the probes contains interesting characteristics when compared to a single detector response, and is often sensitive to the motion of the probes and background curvature \cite{VerSteeg:2007xs, Hu:2015lda, Pozas-Kerstjens:2015gta, Menezes:2017oeb, Koga:2018the, Koga:2019fqh, Cong:2020nec, Zhang:2020xvo, Barman:2021kwg, Barman:2022xht, Barman:2023rhd}, presence of thermal bath \cite{Brown:2013kia, Simidzija:2018ddw, Barman:2021bbw}, spacetime dimensions \cite{Pozas-Kerstjens:2015gta}, and gravitational waves \cite{Xu:2020pbj, Gray:2021dfk, Barman:2023aqk, Barman:2024vah} to be named a few among many others.

Quantum entanglement, being a key feature of quantum theory, plays a crucial role in understanding phenomena at the interface of quantum theory and gravity. In recent years, the study of the role of gravity in quantum information processes such as entanglement and decoherence \cite{K:2023oon,pikovski2015universal}, even at low energy scales compared to the scale at which one requires quantum gravity, has been of interest. These studies hope to gather insights and answer the questions concerning the quantum nature of gravity \cite{bose-etal, marletto-vedral}. In this context, this work provides interesting insights into how the kinematics of quantum probes can combine with the curvature of spacetime to affect the entanglement.

In a recent paper, \cite{K:2023oon}, a rigorous set-up for studying entanglement induced in probes coupled to quantum fields in arbitrary curved spacetime was given which already hinted that spacetime curvature can induce special features in the entanglement between the detectors, and this can have fundamental importance. In this work, we take another step forward and explore the combined role of acceleration and curvature in generating entanglement between the probes. Such an insight becomes particularly relevant since it has already been shown, from single detector analyses \cite{hari:2021gns,K:2023dmj}, that acceleration and curvature can intertwine non-trivially when studying quantum processes in curved spacetime, in a manner that poses tricky challenges for the application of equivalence principle, amongst other things. Introducing two probes, and tracking the entanglement between them, is therefore of particular interest since it might afford a useful way to observe such an intertwining. Indeed, we provide an argument in the last section on how this can be achieved.

\color{black}

\textbf{A quick review of the setup:} We consider two initially uncorrelated two-level Unruh-deWitt detectors to understand the characteristics of induced entanglement due to the background curvature and motion of the detectors. In the present scenario, we consider these detectors to be interacting with a massless scalar field $\phi(x^{i})$ from the background. We denote the two detectors by 1 and 2, with ground and excited states $\l|g_{I}\r>$ and $\l|e_{I}\r>$ and the energy gap $\omega$ between these states for each detector. The interaction Hamiltonian of the system is given by,
\begin{eqnarray}
    \mathcal{H}_{\rm{int}}(\tau) = \sum_{I=1,2}\mu_I \,\chi(\tau_I)\,m_I(\tau_I)\,\phi(x(\tau_I))~.
\end{eqnarray}
Here, $\mu$ is the interaction strength, $\chi$ is the switching function, and $m(\tau)$ is the monopole operator. We consider that the detectors are initially uncorrelated, and thus the product state of the detector-field system in the asymptotic past can be expressed as, $\l|\rm{in}\r>=|0\rangle|g_1\rangle|g_2\rangle$, where $\l|0\r>$ is the vacuum state of the field. In the interaction picture, the state in the asymptotic future will evolve into, $\l|\rm{out}\r>=\mathcal{T}\l\{\exp{(i\int H_{\rm{int}}\,\DM \tau} \l|\rm{in}\r>)\r\}$ with $\mathcal{T}$ used as the time ordering operator. Considering the interaction strength for both the detectors to be the same as $\mu$, and treating it perturbatively one can obtain the reduced density matrix for the detectors as $\rho_{12}={\rm Tr}_{\phi}\, \ket{{\rm out}}\bra{{\rm out}}$. In the bases $\left\{\; \ket{g}_1 \ket{g}_2,\; \ket{e}_1 \ket{g}_2,\; \ket{g}_1 \ket{e}_2,\; \ket{e}_1 \ket{e}_2 \;\right\}$ (Ref. \cite{Koga:2018the}) this reduced density matrix is given by
\begin{eqnarray}\label{eq:density-matrix}
\rho_{12}=
\begin{bmatrix}
1-\mu^2\l(\mathcal{P}_{1}+\mathcal{P}_{2} \r) & 0 & 0 & \mu^2 \mathscr{E}^{*}\\
0 & \mu^2\mathcal{P}_{1} & \mu^2\mathcal{P}_{12} & 0\\
0 & \mu^2\mathcal{P}_{12} & \mu^2\mathcal{P}_{2} & 0\\
\mu^2 \mathscr{E} & 0 & 0 & 0\\
\end{bmatrix} + \mathcal{O}(\mu^4)~.
\end{eqnarray}
For the purpose of understanding entanglement induced between two Unruh-deWitt detectors, we will only need the quantities $\mathcal{P}_{I}$ and $\mathscr{E}$ \cite{Koga:2018the}, which are given by 
\begin{subequations}\label{eq:Pj-E-general}
\begin{eqnarray}
\mathcal{P}_{1} &=& \left \lvert \bra{e} m_{1}(0)\ket{g}_1 \right \rvert^2 \, \mathcal{I}_{1}
\hspace{1cm} ({\rm similarly \; for \; 2})
\\ 
\mathscr{E} &=&  \bra{e} m_{2}(0) \ket{g}_2 \;  \bra{e} m_{1}(0)\ket{g}_1 \, \mathcal{I}_{\varepsilon}~.
\end{eqnarray}
\end{subequations}
The quantities $\mathcal{I}_{I}$ and $\mathcal{I}_{\varepsilon}$ in the previous equation depend on specific detector trajectories, the background spacetime, and also on the specific forms of switching functions. These quantities are given by
\begin{subequations}\label{eq:Ij-Ie-general}
\begin{eqnarray}\label{eq:Ij-general}
\mathcal{I}_{1} &=& 
\int_{-\infty}^{\infty} d\tau^\prime \int_{-\infty}^{\infty}d\tau \;
\chi(\tau) \chi(\tau^{\prime}) 
e^{i \omega (\tau - \tau^{\prime})}
G_{W}(x_{1}(\tau^{\prime}), x_{1}(\tau))
\\ 
i \, \mathcal{I}_{\varepsilon} &=& \int_{-\infty}^{\infty} ds \int_{-\infty}^{\infty}d\tau \; \chi(\tau) \chi(s) 
e^{i \omega (\tau + s)}
G_{F}(x_{2}(s), x_{1}(\tau))~,
\label{eq:Ie-general}
\end{eqnarray}
\end{subequations}
where, $G_{W}(x^{\prime},x)$ and $G_F(x_2,x_1)$ respectively denote the Wightmann function and the Feynman propagator. We should also mention that $\mathcal{P}_{1,2}$ corresponds to individual detector responses, whereas $\mathscr{E}$ is due to the nonlocal correlation between the detectors. As discussed in \cite{Koga:2018the}, the condition for entanglement is $\mathcal{P}_{1} \mathcal{P}_{2} < |\mathscr{E}|^2$, which with the help of the expressions from Eqs. \eqref{eq:Ij-Ie-general} results in $\mathcal{I}_{1} \mathcal{I}_{2} < |\mathcal{I}_{\varepsilon} |^2$.
As elaborated in \cite{Koga:2018the} and \cite{K:2023oon}, we define the measure of the entanglement through Negativity, defined as 
\begin{eqnarray}\label{eq:negativity-general}
    \mathcal{N}(\rho_{12}) &=& c^2_{0}\,\Big[\sqrt{(\mathcal{P}_{1}-\mathcal{P}_{2})^2+4\,|\mathscr{E}|^2}-\mathcal{P}_{1}-\mathcal{P}_{2}\Big]/2 ~.
\end{eqnarray}
With the expression of the monopole moment operator $m(0) = \ket{e}\bra{g} + \ket{g}\bra{e}$ the above definition of Negativity can be simplified to 
\begin{eqnarray}\label{eq:neg}
\mathcal{N} &=& \l[\sqrt{(\mathcal{I}_{1}-\mathcal{I}_{2})^2+4\,|\mathcal{I}_{\epsilon}|^2} - (\mathcal{I}_{1}+\mathcal{I}_{2})\r]/2~,
\end{eqnarray}
where, the definition of \eqref{eq:neg} denotes Negativity upto a factor of $c^2_{0}$. We have not included this factor in Negativity as it denotes a fixed interaction strength and does not carry any information about the background or the motion of the probes. We would also like to mention that when $\mathcal{I}_{1}=\mathcal{I}_{2}=\mathcal{I}$ the above expression of Negativity simplifies to $\mathcal{N} = |\mathcal{I}_{\epsilon}| - \mathcal{I}$.

The manuscript is organized in the following manner. In Sec. \ref{sec:det-trjkt} we start by specifying the accelerated detector trajectories in the de Sitter background. In this section, we also provide the expressions for the geodesic interval, which is a key quantity in Green's functions used to estimate the entanglement-related measures.
In Sec. \ref{sec:Ent-Acltd-probes}, we consider two specific switching scenarios, namely the eternal and finite Gaussian switching, and estimate the Negativity. We indicate our observations on Negativity in this section. In Sec. \ref{sec:Anlys-Maxima} we provide an analysis of the maxima that we will encounter in the Negativity profiles. Finally, we conclude with a discussion of our results in Sec. \ref{sec:concluding-sec}.

\section{Accelerated detectors in de Sitter spacetime}\label{sec:det-trjkt}

In this section, we discuss about detectors in accelerated trajectories in a de Sitter spacetime. We consider a massless conformally coupled scalar field in the background spacetime interacting with these detectors in different trajectories. In particular, we shall delineate these trajectories and find the geodesic interval between two of such probes in these trajectories. We will also provide the expressions of the Wightman function and the Feynman propagator in terms of these geodesic intervals, which will eventually help in discerning the entanglement in the subsequent sections.

\begin{figure}[!h]
  \centering
  \includegraphics[width=15.0cm]{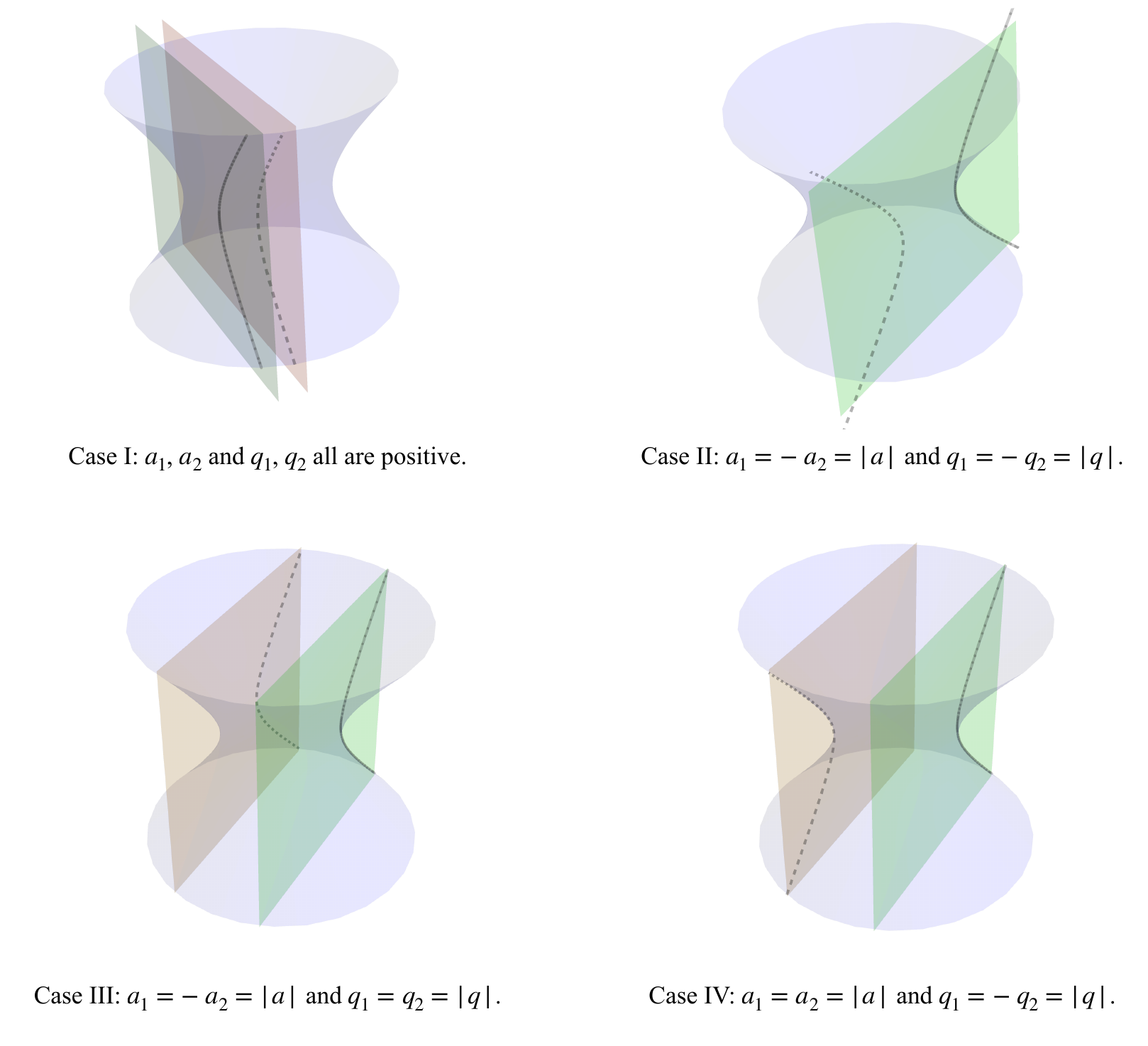}
  \caption{The above figures illustrate the trajectories of uniform acceleration in the embedding space of de Sitter spacetime. The accelerated trajectories are obtained by considering $Y=0=Z$ and $W=\text{constant}$. In the above diagrams, different fixed $W$ are considered, and the curves formed at the interface of these planes and the dS hyperboloid denote the trajectories of uniform acceleration. For instance, in Case I, both the accelerations $a_{1}$ and $a_{2}$ are considered in the same direction with different magnitudes. For Case II, they have the same magnitude but they are in opposite directions, and $q_{1}=-q_{2}$. At the same time, for Case III, the only difference between the trajectories is that the accelerations are in opposite directions with the same magnitude. On the other hand, for Case IV the accelerations are the same and in the same direction with $q_{1}=-q_{2}$, which signifies two $W=\text{constant}$ planes at the opposite sides of the dS hyperboloid.}
  \label{fig:dS-spacetime-embedded}
\end{figure}

The de-Sitter spacetime can be represented as a hyperboloid embedded in a $(4+1)$D Minkowski spacetime (see Ref. \cite{weinberg1972gravitation,deser1997accelerated}) using the embedding equation
\begin{equation}
  -T^{2}+ X^{2} + Y^{2} + Z^{2} + W^{2} = \frac{1}{\Lambda},
\end{equation}
where $\Lambda$ is the constant positive curvature ($[\Lambda] = [L]^{-2}$) for dS spacetime and $(T,X,Y,Z,W)$ are the embedding coordinates. The visualization of the dS spacetime is illustrated in Fig. \ref{fig:dS-spacetime-embedded} in terms of the embedding coordinates, where $y$ and $z$ coordinates are suppressed. The spacetime is characterized by a constant Minkowski norm surface, much like a ``sphere" of constant radius in $3-$dimensions \cite{weinberg1972gravitation}. 

Let us consider 2 uniformly accelerated trajectories in dS spacetime which will correspond to Rindler trajectories in embedding $(4+1)$D spacetime with $5-$acceleration $q_{1}$ and $q_{2}$, where $q = \sqrt{a^{2}+\Lambda}$. For a suitable choice of initial conditions ($y=z=0$), the trajectories will lie in $W=\text{constant}$ planes as shown in Fig. \ref{fig:dS-spacetime-embedded}. The proper time parametrization of the trajectories (with zero initial relative velocity) in $(4+1)$D flat spacetimes is given by (see Ref. \cite{hari:2021gns}), 
\begin{subequations}\label{eq:tau-para-AccTrj}
\begin{eqnarray}
    Z^{A}_{1}(\tau_{1}) &=& \left( \frac{\sinh(q_{1}\tau_{1})}{q_{1}}, \frac{\cosh(q_{1}\tau_{1})}{q_{1}},0,0,\frac{a_{1}}{q_{1}\sqrt{\Lambda}} \right)\\
    Z^{A}_{2}(\tau_{2}) &=& \left( \frac{\sinh(q_{2}\tau_{2})}{q_{2}}, \frac{\cosh(q_{2}\tau_{2})}{q_{2}},0,0,\frac{a_{2}}{q_{2}\sqrt{\Lambda}} \right)~,
\end{eqnarray}
\end{subequations}
where the subscript $1$ and $2$ correspond to two different detector trajectories. 

It is clear from the expression for Negativity that, we will require the Wightman function and the Feynman propagator to evaluate the Negativity. We will consider conformally coupled massless scalar field in de Sitter background. The expression for the Wightman function and Feynman propagator are, respectively (see Refs. \cite{Allen:1985ux,Garbrecht:2004du}),
\begin{subequations}\label{eq:gen-Gw-Gf}
\begin{eqnarray}\label{eq:gen-Wightman}
    \mathcal{G}_{W}(x,x^\prime) &=& -\frac{1}{4\pi^{2}}\l[\frac{\Lambda}{4\sin^2\l(\sqrt{\Lambda}\sigma^2_{\epsilon}(x,x^\prime)/2\r)}\r]~,\\
    \mathcal{G}_{F}(x,x^\prime) &=& \frac{i}{4\pi^{2}}\l[\frac{\Lambda}{4\l(\sin^2\l(\sqrt{\Lambda}\sigma^2(x,x^\prime)/2\r)+i\epsilon\r)}\r]~.\label{eq:gen-Feynman}
\end{eqnarray}
\end{subequations}
where, $\sigma^2(x,x^\prime)$ is the square of geodesic interval between the spacetime points $x$ and $x^\prime$. For the generation of entanglement between the detectors, we need to consider the interplay between the local noise and nonlocal correlation. The local noise (detector response) is estimated using the Wightman function and the nonlocal correlations are evaluated using the Feynman propagator. For local noise, the pullback of the Wightman function onto the trajectory is used, which requires the geodesic interval between two points on a single detector trajectory. For the nonlocal correlator, the geodesic interval between two points on two different detector trajectories should be utilized in the Feynman propagator.

Since the hyperboloid of dS spacetime is a surface of constant ``radius" $(1/{\sqrt{\Lambda}})$, the corresponding arc length (geodesic distance in de Sitter) between two points is related to the chord length (geodesic distance in higher dimensional Minkowski) by the equation (see Ref. \cite{Garbrecht:2004du}),
\begin{equation}\label{eq:sgm-geod}
   \sigma_{\text{geod}}(x,x^{\prime}) = \frac{2}{\sqrt{\Lambda}} \sin^{-1}\left( \frac{\sqrt{\Lambda}\Sigma(x,x^{\prime})}{2} \right)~.
\end{equation}
Here, $\sigma_{\text{geod}}(x,x^{\prime})$ is the geodesic interval in de Sitter spacetime between the point $x$ and $x^{\prime}$ and $\Sigma(x,x^\prime)$ is the geodesic interval between the same points in the higher dimensional Minkowski spacetime.

It is straightforward to compute the Minkowski distance between two spacetime points on an accelerated trajectory as,
\begin{equation}
    \Sigma^{2} = \eta_{A B}\left(Z^A(\tau)- Z^A(\tau^\prime)\right)\left(Z^B(\tau)- Z^B(\tau^\prime)\right) = -\frac{4}{q^2}\sinh^2{\left( \frac{q (\tau-\tau^\prime)}{2} \right)}~,
    \label{eq:chordal-general}
\end{equation}
which gives us the chordal distance between two points on a single trajectory. Similarly, for two spacetime points on two different accelerated trajectories is given by,
\begin{equation}\label{eq:chordal-general}
    \Sigma^{2} = \eta_{A B}(Z^{A}_{2}(\tau_{2})- Z^{A}_{1}(\tau_{1}))(Z^{B}_{2}(\tau_{2}) - Z^{B}_{1}(\tau_{1})) = \left( \frac{2}{\Lambda} - \frac{2a_{1}a_{2}}{\Lambda q_{1}q_{2}} -\frac{2 \cosh(q_{2}\tau_{2}-q_{1}\tau_{1})}{q_{1}q_{2}} \right)~,
\end{equation}
which gives us the chordal distance between two points on the respective trajectories. 
The above equation for geodesic distance reduces to that of Minkowski for $\Lambda \to 0$, i.e.
\begin{equation}\label{eq:SigMink-Feynmn}
   { \sigma^{2}}|_{\text{dS}} = {\sigma^{2}}|_{\mathcal{M}} + \mathcal{O}(\sqrt{\Lambda})\,,~~
\textup{where}~~
    {\sigma^{2}}|_{\mathcal{M}} = \frac{1}{a_{1}^{2}} + \frac{1}{a_{2}^{2}} - \frac{2 \cosh(a_{2}\tau_{2}-a_{1}\tau_{1})}{a_{1}a_{2}}~.
\end{equation}
\color{black}

%%%%%%%%%%%%%%%%%%%%%%%%%%%%%%%%%%%%%%%%%%%%%%%%%%%%%%%%%%%%%%%%%%%%%%%

\begin{figure}[!h]
  \centering
  \includegraphics[width=7.0cm]{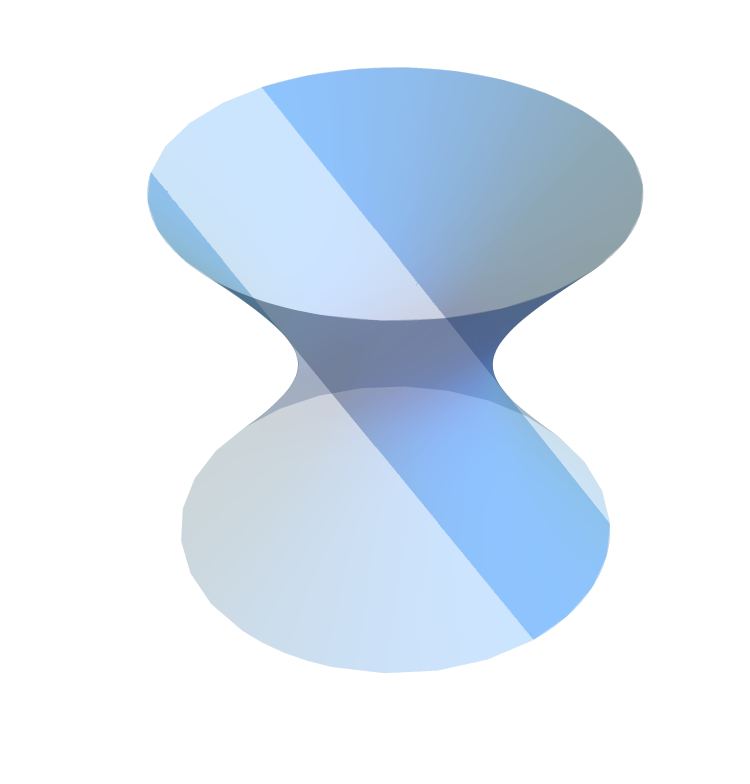}
  \hskip 20pt
  \includegraphics[width=7.0cm]{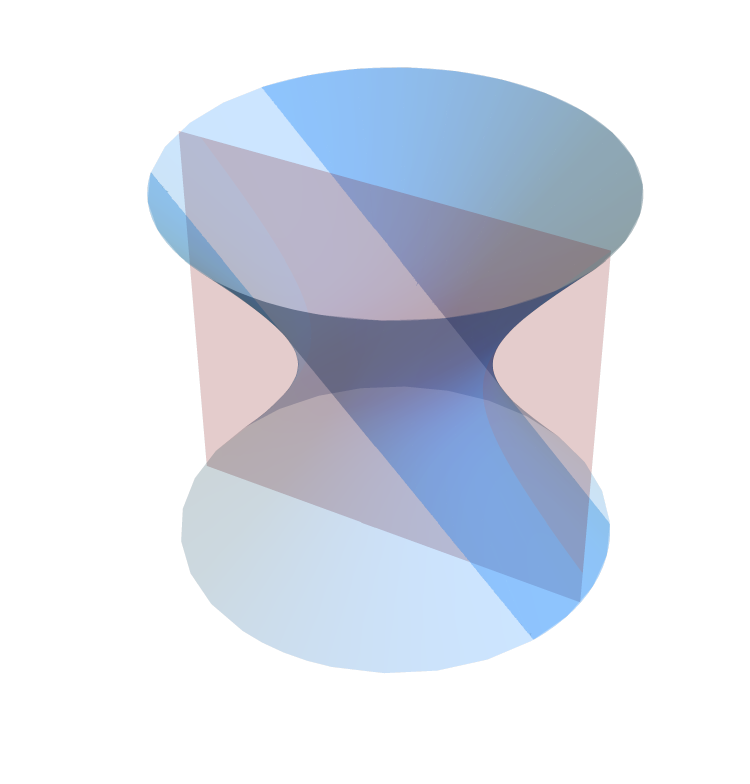}
  \caption{In the above figure we have illustrated the flat cosmological patch in a de Sitter hyperboloid. In both of the above diagrams, the shaded region in darker blue signifies the region of a flat cosmological patch. In particular, in the left figure, this cosmological patch is illustrated separately. It should be noted that a flat cosmological patch covers only half of the de Sitter hyperboloid, see \cite{Hawking_Ellis_1973}. At the same time, in the right diagram, we have depicted an additional $W=\text{constant}$ surface in a transparent red colour. We notice that for this particular $W=\text{constant}$ surface only one accelerated trajectory (these trajectories are obtained at the intersection of this surface and the de Sitter hyperboloid) lies in the flat cosmological patch. We would also like to mention that the specific detector trajectories of Case II \& IV resemble the scenario of the right diagram, where one trajectory resides inside a flat cosmological patch and the other one is outside.}
  \label{fig:DS-flatFLRW}
\end{figure}
%%%%%%%%%%%%%%%%%%%%%%%%%%%%%%%%%%%%%%%%%%%%%%%%%%%%%%%%%%%%%%%%%%%%%%%

Let us now discuss a few accelerated trajectories in the dS background that are straightforward to visualize. We will be interested in certain pairs of trajectories as indicated below:
\begin{enumerate}
    \item \textbf{Case I}: The trajectories are in the same static patch, i.e., they are in the same causal patch, with different magnitudes of acceleration. The chordal distance between the two detectors, in this case, can be obtained from \eqref{eq:chordal-general}. For a schematic diagram of the trajectories in a de Sitter hyperboloid, see Fig. \ref{fig:dS-spacetime-embedded}.
    
    \item \textbf{Case II}: The trajectories are in different static patches with the same magnitudes of acceleration. Both $4-$acceleration $a$, and $5-$acceleration $q$ are opposite, $a_1=-a_2=|a|;\,\,q_1=-q_2=|q|$. In this case, \eqref{eq:chordal-general} reduces to,
    \begin{eqnarray}\label{eq:C2-geodesic-dist}
        \Sigma^2=\frac{4}{q^2}\cosh^2\l(\frac{q}{2}(\tau_2+\tau_1)\r)
    \end{eqnarray}
    The above equation is the same as the geodesic interval between two trajectories in opposite Rindler wedges with the same magnitude of acceleration. For a schematic diagram see Fig. \ref{fig:dS-spacetime-embedded}. One can also notice that if one of these two accelerated trajectories is in a flat cosmological patch the other will be outside, see Fig. \ref{fig:DS-flatFLRW}. 
    
    \item \textbf{Case III}: The trajectories are in the same static patch with the same magnitudes of acceleration but are oriented in $\theta = 0$ and $\theta=\pi$ planes respectively of de Sitter spacetime. The 4 and 5 accelerations are related by, $a_1=-a_2=a;\,\,q_1=q_2=q$. Equation \ref{eq:chordal-general} simplifies to,
    \begin{eqnarray}\label{eq:C3-geodesic-dist}
        \Sigma^2=\frac{2}{\Lambda q^2}\l[q^2+a^2-\Lambda\,\cosh\l(q(\tau_2-\tau_1)\r)\r]
    \end{eqnarray}
    
    \item \textbf{Case IV}: The trajectories are in different static patches with the same magnitude of acceleration but are oriented in $\theta = 0$ and $\theta=\pi$ planes respectively of de Sitter spacetime. The 4 and 5 accelerations are related by, $a_1=a_2=a;\,\,q_1=-q_2=q$. Equation \ref{eq:chordal-general} simplifies to,
    \begin{eqnarray}\label{eq:C4-geodesic-dist}
        \Sigma^2=\frac{2}{\Lambda q^2}\l[q^2+a^2+\Lambda\,\cosh\l(q(\tau_2+\tau_1)\r)\r]~.
    \end{eqnarray}

    similar to Case II, here also if one of the trajectories is in the flat cosmological patch the other will be outside, see Fig. \ref{fig:DS-flatFLRW}.
\end{enumerate}

We want to mention that we have explicitly depicted these detector trajectories through schematic diagrams in Fig. \ref{fig:dS-spacetime-embedded}. We will be using the expression of $\Sigma^2$ obtained here in the next section, where we estimate the entanglement between the detectors.

%%%%%%%%%%%%%%%%%%%%%%%%%%%%%%%%%%%%%%%%%%%%%%%%%%%%%%%%%%%%%%%%%%%%%%%%%%%%%%%%%%%%%%%%%%%%%%%%%

\section{Entanglement between accelerated detectors}\label{sec:Ent-Acltd-probes}

In this section, we estimate the entanglement, especially the measure of it in terms of Negativity. In this regard, we consider two specific scenarios of switching functions. First, we consider the eternal interaction between the detectors and the field, which is given by $\chi(\tau)=1$. With this switching the response functions and the Negativity will remain impervious to transient effects, and thus is useful in understanding the generic role of motion and background in Negativity. Second, we consider finite switching in terms of the Gaussian function $\chi(\tau) = e^{-{\tau^{2}}/{T^{2}}}$ to estimate entanglement. This kind of finite switching is relevant from the experimental point of view as practically one cannot carry out an experiment for infinite time.

\subsection{Eternal switching}

In this part, we consider eternal switching, i.e., $\chi(\tau)=1$, between the detectors and the background scalar field. With this consideration, we estimate the quantity $\mathcal{I}_{1}$, which signifies the individual detector transition probabilities and the quantity $\mathcal{I}_{\varepsilon}$, corresponding to the nonlocal entangling term, and the Negativity constructed out of the previous two terms. We would like to mention that we will find all these quantities to be divergent. However, the time rates of these quantities, which for eternal switching is obtained by dividing by infinite time, will be finite. We define these rates to be $\mathcal{R}_{1}$ and $\mathcal{R}_{\varepsilon}$ that correspond to $\mathcal{I}_{1}$ and $\mathcal{I}_{\varepsilon}$ respectively.

\subsubsection{Evaluation of the local terms in Negativity}

Here we evaluate the local terms $\mathcal{I}_{1}$ with the eternal switching $\chi(\tau)$. In this regard, we consider the Wightman function for detectors in accelerated trajectories in a de Sitter background as obtained from Eq. \eqref{eq:gen-Wightman} using Eqs. \eqref{eq:sgm-geod} and \eqref{eq:chordal-general}. We utilize this Wightman function in Eq. (\ref{eq:Ij-general}) to find the local terms $\mathcal{I}_{1}$. To evaluate the expression of $\mathcal{I}_{1}$ we make use of a change of variables $u=\tau_{1}-\tau'_{1}$ and $\tau=\tau_{1}$. We have provided the explicit evaluation of $\mathcal{I}_{1}$ in Appendix \ref{Appn:EternalSwtch-Ije}, Eq. \eqref{AppEq:Keq1-Ij-1}. It is to be noted that the subscript $1$ corresponds to the first detector, and this local term for the second detector can be evaluated in a similar manner. In particular, these local terms correspond to the individual detector transition probabilities, and we define the rate of these transition probabilities as $\mathcal{R}_{1}=\mathcal{I}_{1}/\l\{\lim_{\mathbb{T}\to\infty}\int_{-\mathbb{T}}^{\mathbb{T}}d\tau\r\}$. With the help of expression of $\mathcal{I}_{1}$, the expression of $\mathcal{R}_{j}$ becomes
\begin{eqnarray}\label{eq:Keq1-Rj-1}
    \mathcal{R}_{1} = \frac{\omega}{\pi}\,\frac{1}{e^{2 \pi  \omega/q_{1}-1}}~.
\end{eqnarray}
This probability rate for the second detector can be obtained in a similar manner. We would like to mention that in the above definition, the rate for detector $1$ is defined in terms of $\tau=\tau_{1}$, whereas for detector $2$ it is defined in terms of $\tau=\tau_{2}$. This would lead to an issue of uniformity when defining the Negativity rate. However, one can also notice (from \eqref{eq:tau-para-AccTrj}) that when the magnitudes of the five accelerations $(q)$ in both of the cases are the same, i.e. when $q_{1}=q_{2}$, the proper time parametrizations for the two different trajectories are essentially done by the same parameter $\tau_{1}=\tau_{2}=\tau$. In our subsequent analysis, we will observe that in all our considered trajectories the magnitudes of these five accelerations are the same for both of these detectors, and it will not cause any issue in defining the Negativity rate.

\subsubsection{Evaluation of the non-local term in Negativity}

\begin{itemize}
     \item \textbf{Case I:} To evaluate the rate of the nonlocal term $\mathcal{R}_{\varepsilon}$ corresponding to the eternal switching and the detector trajectories of Case I from Sec. \ref{sec:Ent-Acltd-probes}, we consider the expression of the Feynman propagator from Eq. \eqref{eq:gen-Feynman}. Moreover, to evaluate the nonlocal integral $\mathcal{I}_{\varepsilon}$ we consider a change of variables $v=\tau_{1}+\tau_{2}$ and $\tau=\tau_{1}$. The explicit evaluation of this integral is given in Appendix \ref{Appn:EternalSwtch-Ije}, see Eq. \eqref{AppEq:C1-Keq1-Ie-3}. The rate of this nonlocal term is defined as $\mathcal{R}_{\varepsilon}=\mathcal{I}_{\varepsilon}/\l\{\lim_{\mathbb{T}\to\infty}\int_{-\mathbb{T}}^{\mathbb{T}}d\tau\r\}$. This rate can be expressed as (with the help of Eq. \eqref{AppEq:C1-Keq1-Ie-3})
\begin{eqnarray}\label{eq:C1-Keq1-Re}
    \mathcal{R}_{\varepsilon} &=& \frac{\Lambda  q_{1} \left\{\coth \left(\frac{\pi  \omega}{q_{2}}\right)-1\right\} e^{\frac{\omega \,\pi}{q_{2}}} \sinh \left[\frac{\omega \left(\pi -i \cosh ^{-1}\left(\frac{q_{1} q_{2}-a_{1} a_{2}}{\Lambda }\right)\right)}{q_{2}}\right]}{2  \sqrt{(a_{1} a_{2}+\Lambda -q_{1} q_{2})(a_{1} a_{2}-\Lambda -q_{1} q_{2})}}\,\frac{\delta( \omega\, (q_{1}+q_{2})/q_{2})}{\l\{\lim_{\mathbb{T}\to\infty}\int_{-\mathbb{T}}^{\mathbb{T}}d\tau\r\}}=0~,
\end{eqnarray}
where $\delta(x)$ denotes the Dirac delta distribution. For the above scenario, the Dirac delta gives zero as both $q_{1}$ and $q_{2}$ are positive real for the trajectories of Case I.
\vspace{0.2cm}

\item \textbf{Case II:} Here we consider the Feynman propagator from Eq. \eqref{eq:gen-Feynman} with the trajectories of Case II from Eq. \eqref{eq:C2-geodesic-dist}. Here also we consider a change of variables to $v=\tau_{1}+\tau_{2}$ and $\tau=\tau_{1}$, to obtain the integral $\mathcal{I}_{\varepsilon}$. In Appendix \ref{Appn:EternalSwtch-Ije}, Eqs.
\eqref{AppEq:C2-Keq1-Ie-1}-\eqref{AppEq:C2-Keq1-Ie-3}, we have provided the explicit evaluation of this integral and rate $\mathcal{R}_{\varepsilon}$. In particular, the explicit expression of this rate in this scenario is 
\begin{eqnarray}\label{eq:C2-Keq1-Re}
    \mathcal{R}_{\varepsilon} &=& \frac{i \, \omega \, e^{\pi  \omega/q}}{2 \, \pi \, \left(e^{2 \pi  \omega/q}-1\right)}~.
\end{eqnarray}
Here we would like to point out a few features of this rate for nonlocal entangling terms. From the above expression, we see that the quantity $\mathcal{R}_{\varepsilon}$ is a function of $q$ as a whole, and we know $q=\sqrt{a^2+\Lambda}$. Therefore, with respect to the changes in $a$ and $\Lambda$ the qualitative changes in $\mathcal{R}_{\varepsilon}$ will be similar. For instance, when the value of $a$ or $\Lambda$ increases the modulus of $\mathcal{R}_{\varepsilon}$ increases, and will eventually result in an infinite value as $q\to \infty$. One can also observe that in the limit of $q\to 0$, i.e., for static detectors in the Minkowski background, the rate of the nonlocal term vanishes indicating a vanishing entanglement between the detectors.
\vspace{0.2cm}

\item \textbf{Case III:} To evaluate the rate of the nonlocal term for Case III we consider the geodesic interval for this scenario as provided in Eq. (\ref{eq:C3-geodesic-dist}). We utilize it in the Feynman propagator of Eq. \eqref{eq:gen-Feynman}. We observed that in this scenario also it is easier to evaluate $\mathcal{I}_{\varepsilon}$ with a change of variables to $v=\tau_{1}+\tau_{2}$ and $\tau=\tau_{1}$. The explicit evaluation is provided in Eqs. \eqref{AppEq:C3-Keq1-Ie-1}-\eqref{AppEq:C3-Keq1-Ie-3} of Appendix \ref{Appn:EternalSwtch-Ije}. In particular, the rate of this nonlocal term $\mathcal{I}_{\varepsilon}$ has the expression
\begin{eqnarray}\label{eq:C3-Keq1-Re}
    \mathcal{R}_{\varepsilon} &=& -\frac{\Lambda\,q  \,\text{csch}\left(\frac{\pi\,\omega}{q}\right) \sinh \left[\omega \left\{\pi -i \cosh ^{-1}\left(\frac{a^2+q^2}{\Lambda }\right)\right\}/q\right]}{4\,  \sqrt{(a^2-\Lambda +q^2)\left(a^2+\Lambda +q^2\right) }}~\frac{\delta(2\,\omega)}{\l\{\lim_{\mathbb{T}\to\infty}\int_{-\mathbb{T}}^{\mathbb{T}}d\tau\r\}}=0~.
\end{eqnarray}
In the previous expression $\delta(x)$ denotes a Dirac delta distribution. The above expression vanishes as we have considered a positive detector energy gap $\omega>0$. This signifies that for Case III the rate of the nonlocal entangling term is zero for eternal switching. Therefore, like Case I, in Case III also the entanglement there is no entanglement.\vspace{0.2cm}

\item \textbf{Case IV:} For Case IV we consider the expression of the geodesic interval from Eq. (\ref{eq:C4-geodesic-dist}) and utilize it in the expression of the Feynman propagator of Eq. \eqref{eq:gen-Feynman}. here also we consider a change of variables to $v=\tau_{1}+\tau_{2}$ and $\tau=\tau_{1}$. The explicit evaluation of the integral $\mathcal{I}_{\varepsilon}$ is provided in Appendix \ref{Appn:EternalSwtch-Ije}, see Eqs. \eqref{AppEq:C4-Keq1-Ie-1} and \eqref{AppEq:C4-Keq1-Ie-2}. In particular, we obtain the rate of the nonlocal term as
\begin{eqnarray}\label{eq:C4-Keq1-Re}
    \mathcal{R}_{\varepsilon} &=& \frac{\Lambda\,q\,e^{-\frac{i \omega \cosh ^{-1}\left(-\frac{a^2+q^2}{\Lambda }\right)}{q}} \left[-1+e^{2 \omega \left\{\pi +i \cosh ^{-1}\left(-\frac{a^2+q^2}{\Lambda }\right)\right\}/q}\right]}{4 \pi  \left(e^{\frac{2 \pi  \omega}{q}}-1\right) \sqrt{\left(a^2-\Lambda +q^2\right)\left(a^2+\Lambda +q^2\right)}}~.
\end{eqnarray}
Here we would like to mention that, unlike the expression of \eqref{AppEq:C2-Keq1-Ie-3} the above $\mathcal{R}_{\varepsilon}$ is not a function of $q$ as a whole. Therefore, its behaviour could be different as $a$ or $\Lambda$ changes, and we shall observe that this is indeed the case. For instance, for very large four acceleration the rate of the nonlocal term behaves as $\mathcal{R}_{\varepsilon}= (\Lambda/8 \pi ^2 a)  \left\{\pi +i\, \log \left(-4 a^2/\Lambda \right)\right\}+\mathcal{O}(1/a^2)$, which vanishes as $a\to \infty$. Whereas, for large curvature the same quantity behaves as $\mathcal{R}_{\varepsilon}= i \, \sqrt{\Lambda }/4 \pi ^2 + \mathcal{O}(1/\sqrt{\Lambda})$, i.e., it keeps increasing as $\Lambda$ increases. When $a\to 0$ the rate of the nonlocal term reduces to \eqref{AppEq:C2-Keq1-Ie-3} with $a=0$, i.e., it becomes $\mathcal{R}_{\varepsilon}= i\, \omega\, e^{\frac{\pi  \omega}{\sqrt{\Lambda }}}/\l\{2 \pi  \left(e^{\frac{2 \pi  \omega}{\sqrt{\Lambda }}}-1\right)\r\}$. On the other hand, a similar limit for $\Lambda\to 0$ results in a diverging $\mathcal{R}_{\varepsilon}$, indicating that this limit is unphysical.\\

We shall use these expressions of the nonlocal entangling terms $\mathcal{R}_{\varepsilon}$ from Eqs. (\ref{AppEq:C1-Keq1-Ie-3})-(\ref{AppEq:C4-Keq1-Ie-2}) along with the term $\mathcal{R}_{j}$ from Eq. (\ref{eq:Keq1-Rj-1}) to obtain the Negativity rate in the next part.

\end{itemize}

%%%%%%%%%%%%%%%%%%%%%%%%%%%%%%%%%%%%%%%%%%%%%%%%%%%%%%%%%%%%%%%%%%%%%%%%%%%%%%%%%%%%%%%%%%%%%%%%%
\begin{figure}[h!]
\centering
\includegraphics[width=8.0cm]{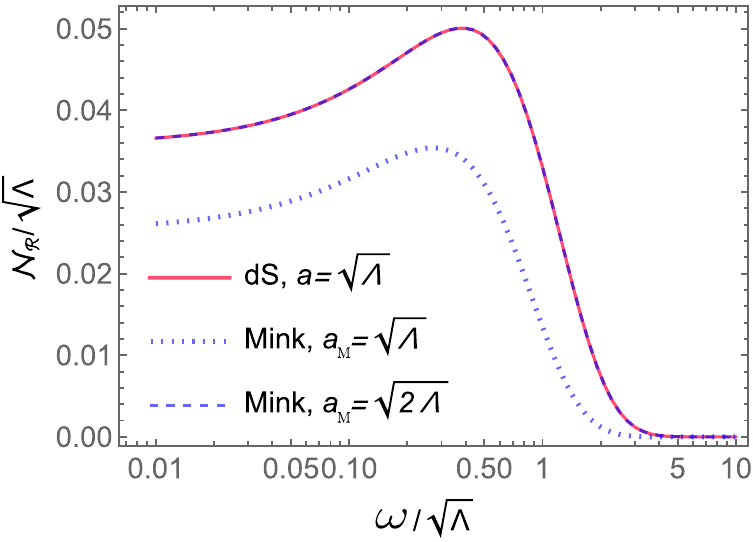}
\caption{\textbf{(Eternal switching) Case II:} In the above figure we have plotted the rate of Negativity for Case II as a function of the detector energy gap corresponding to the eternal switching scenario $\chi(\tau)=1$. From this plot, we observe that this rate is the same as the Negativity rate of two accelerated probes in anti-parallel motion in the Minkowski background with acceleration $(a^2+\Lambda)^{1/2}$, which is also evident from the expression of \eqref{eq:C2-Keq1-Re}.}
\label{fig:NR-eternal-caseII}
\end{figure}

\begin{figure}[h!]
\centering
\includegraphics[width=12.0cm]{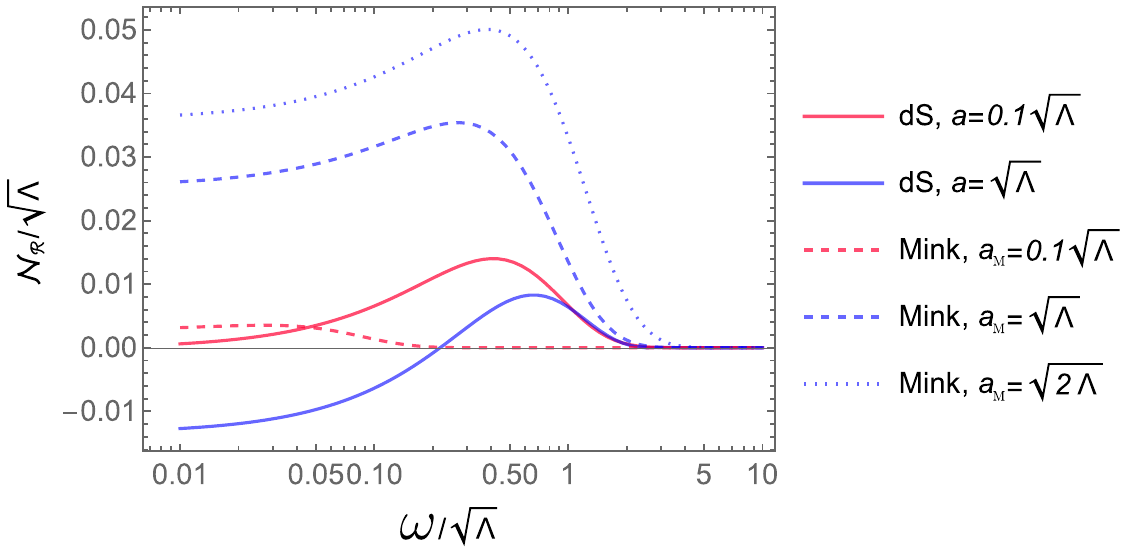}
\caption{\textbf{(Eternal switching) Case IV:} In the above figure we have plotted the Negativity rates corresponding to the detector trajectories of Case IV as functions of the dimensionless energy gap. We have also plotted the Minkowski Negativity rate in the same plot. The expression of the de Sitter Negativity rate is obtained utilizing \eqref{eq:C4-Keq1-Re}. It should be noted that in the Minkowski background, the acceleration is considered to be $a_{M}=(a^2+\Lambda)^{1/2}$.}
\label{fig:NR-eternal-caseIV}
\end{figure}

\iffalse
\begin{figure}[h!]
\centering
\includegraphics[width=8.5cm]{Figs/Case-4-scaled.pdf}
\caption{\textbf{Eternal switching, Case IV:} In the above plots we provide the schematic diagram for detector trajectories mentioned in Case IV (with $q_1=q_2=q$ and $a_1=-a_2=a$), and plot the corresponding rate of Negativity. On the right, we have plotted the Negativity rates for detectors in accelerated motion respectively in the de Sitter and Minkowski backgrounds for eternal switching. The expression of the de Sitter Negativity rate is taken from \eqref{eq:C4-Keq1-Ie-2}. It should be noted that in the Minkowski background, the acceleration is considered to be $(\Lambda+a^2)^{1/2}$.}
\label{fig:NR-eternal-caseIV}
\end{figure}\fi
%%%%%%%%%%%%%%%%%%%%%%%%%%%%%%%%%%%%%%%%%%%%%%%%%%%%%%%%%%%%%%%%%%%%%%%%%%%%%%%%%%%%%%%%%%%%%%%%%

\subsubsection{Observations on the rate of Negativity}

Here we express the Negativity, more precisely the Negativity rates, corresponding to different accelerated detector trajectories in the de Sitter background for eternal switching, i.e., $\chi(\tau)=1$. Taking the expression of Negativity from \eqref{eq:neg}, we define the rate of Negativity for eternal switching as, $\mathcal{N}_{\mathcal{R}}=\mathcal{N}/\l(\lim_{\mathbb{T}\to\infty}\,\int_{-\mathbb{T}}^{\mathbb{T}}d\tau\r)$. Then in terms of local and nonlocal terms, this rate is given by
\begin{eqnarray}\label{eq:Neg-rate-eternal}
    \mathcal{N}_{\mathcal{R}} &=& \l[\sqrt{(\mathcal{R}_{1}-\mathcal{R}_{2})^2+4\,|\mathcal{R}_{\epsilon}|^2} - (\mathcal{R}_{1}+\mathcal{R}_{2})\r]/2~.
\end{eqnarray}
Here, the expressions of $\mathcal{R}_{j}$ are obtained from \eqref{eq:Keq1-Rj-1}, while the expressions of $\mathcal{R}_{\varepsilon}$ are obtained from Eqs. \eqref{eq:C1-Keq1-Re}-\eqref{eq:C4-Keq1-Re}. From these equations, one can observe that for Case I and Case III the rate $\mathcal{R}_{\varepsilon}$ of the nonlocal term vanishes for eternal switching. Therefore, in these cases, the Negativity will be $\mathcal{N}_{\mathcal{R}} = \l\{|\mathcal{R}_{1}-\mathcal{R}_{2}| - (\mathcal{R}_{1}+\mathcal{R}_{2})\r\}/2$, which is bound to be negative as both $\mathcal{R}_{j}$ are positive and real. Therefore, \textit{there is no entanglement generation in these two cases (Case I and III) with eternal switching}. Non-vanishing entanglement is possible only when $\mathcal{R}_{\varepsilon}$ is non-zero, which can be obtained for Case II and Case IV from Eqs. \eqref{eq:C2-Keq1-Re} and \eqref{eq:C4-Keq1-Re}.

We have plotted the Negativity rates for Case II and Case IV in Figs. \ref{fig:NR-eternal-caseII} and \ref{fig:NR-eternal-caseIV} respectively. In Fig. \ref{fig:dS-spacetime-embedded}, we have depicted the accelerated detector trajectories corresponding to these different cases through schematic diagrams. The main observations from these figures are as follows.
\begin{itemize}
    \item From Fig. \ref{fig:NR-eternal-caseII} depicting Case II, we can see that the Minkowski and de Sitter Negativity rates are related through the mapping, $a^{2}\rightarrow a^{2}+\Lambda$, which is also evident from the expressions of the de Sitter local and nonlocal terms of \eqref{eq:Keq1-Rj-1} and \eqref{eq:C2-Keq1-Re}. In Fig. \ref{fig:NR-eternal-caseII}, this fact can be visualized through the red-solid (de Sitter with $a=\sqrt{\Lambda}$) and blue-dashed (Minkowski with $a=\sqrt{2\,\Lambda}$) curves, which completely overlap. From Fig. \ref{fig:NR-eternal-caseIV}, we observe that a similar analogy between the Minkowski and de Sitter Negativity rates is not possible for Case IV, which is also supported by the expression of the de Sitter nonlocal term of Eq. \eqref{eq:C4-Keq1-Re}.

    \item In both Figs. \ref{fig:NR-eternal-caseII} and \ref{fig:NR-eternal-caseIV} we observe that the Negativity rates do not change monotonically with the detector energy gap. At first, it increases with the increasing energy gap and then keeps decreasing with a further increase in energy. We believe this behaviour is a feature specific to the presence of acceleration and in Case II to the presence of curvature.

    \item From Fig. \ref{fig:NR-eternal-caseII}, we observe that the Negativity is always positive and becomes infinitesimal for a large energy gap $\omega$. Furthermore, Negativity increases with increasing acceleration and curvature. From Fig. \ref{fig:NR-eternal-caseIV} we observe that Negativity can also be negative, which corresponds to vanishing entanglement, for small energy gaps and high acceleration. However, with increasing curvature $\Lambda$ this situation may change, see Fig. \ref{fig:NR-eternal-caseIV}. Therefore, from these plots, one can assert that \textbf{\emph{presence of curvature enhances entanglement for accelerated probes}}.
    
    \item On the other hand, acceleration itself can have an enhancing or inhibiting effect on the entanglement. For instance, in Case II increasing acceleration enhances the entanglement, and in Case IV it inhibits the entanglement, see Figs. \ref{fig:NR-eternal-caseII} and \ref{fig:NR-eternal-caseIV}. This assertion is also strengthened by the fact that for parallel acceleration of the two probes of Case I there is no entanglement even in the Minkowski background. Therefore, for accelerated detectors in a de Sitter spacetime, \textbf{\emph{the entanglement has disparate features w.r.t. acceleration as compared to single detector responses}}, see Refs. \cite{hari:2021gns, deser1997accelerated}.

    \item From Fig. \ref{fig:NR-eternal-caseIV} one also observes that for low detector energy gap, there could be vanishing entanglement in de Sitter spacetime with small values of $\Lambda$, and the Negativity becomes positive only above certain values of energy gap $\omega$ as we increase $\Lambda$. We would also like to mention that even though Case IV has some unique features in de Sitter spacetime, the detector trajectories belong to different cosmological patches and thus constructing the experimental set-up may not be possible.
    
\end{itemize}

\subsection{Finite Gaussian switching}\label{subsec:Gauss-negativity}
In this part, we consider finite switching described by the Gaussian function $\chi(\tau) = e^{-{\tau^{2}}/{T^{2}}}$, for the interaction between the detectors and the background field. Subsequently, we will be using this switching function to calculate the local terms $\mathcal{I}_{1}$, the nonlocal terms $\mathcal{I}_{\epsilon}$, and the Negativity $\mathcal{N}$.

\subsubsection{Evaluation of the local terms in Negativity}

For the Gaussian switching, we first try to evaluate the local terms $\mathcal{I}_{1}$ as specified in Eq. \eqref{eq:Ij-general}. In this regard, a suitable way forward is to consider a change of variables to $u=\tau_{1}-\tau'_{1}$ and $\tau=\tau_{1}$. Here we should mention that in both Minkowski and de Sitter backgrounds if the detectors are accelerated the previous considerations will simplify the calculations. However, even with this, we will not get a final analytical expression for the local terms. According to our understanding, we will have to deal with minimum numerical computations when one of the time integrals in $\mathcal{I}_{1}$ is carried out analytically. In that scenario, this integral for the de Sitter background looks like
\begin{equation}\label{eq:KeqG-IjDS}
    \mathcal{I}_{1} = \frac{T^{2}}{4 \pi}\int_{-\infty}^{\infty} dk \frac{e^{\frac{2k\pi}{\sqrt{a^{2}+ \Lambda}} - \frac{k^{2}T^{2}}{2} + (\omega - k)\epsilon}}{e^{\frac{2 k \pi}{\sqrt{a^{2}+\Lambda}}} - e^{\frac{2 \pi \omega}{\sqrt{a^{2}+\Lambda}}}}(k - \omega)~,
\end{equation}
where $k$ is some intermediate variable we have converted our integral into. The explicit procedure to obtain this form of $\mathcal{I}_{1}$ is given in Appendix \ref{Appn:GaussSwtch-Ije}. From the above expression, it is easy to obtain the expression of $\mathcal{I}_{1}$ corresponding to an accelerated observer in the Minkowski background by putting $\Lambda=0$. We would also like to mention that if the detectors were inertial it would have been possible to obtain analytical expressions for the local terms, as also done in \cite{LSriramkumar_1996}.

\subsubsection{Evaluation of the non-local term in Negativity}

Before we proceed to evaluate the nonlocal entangling term $\mathcal{I}_{\varepsilon}$, we would like to recall that there were four different pairs of detector trajectories that we considered in Sec. \ref{sec:det-trjkt}. These trajectories are illustrated in Fig. \ref{fig:dS-spacetime-embedded}. We shall consider each of these cases corresponding to the different pair of trajectories individually to evaluate the nonlocal term. 
\begin{itemize}
    \item \textbf{Case I:} Here we consider the situation for Case I, as described in Sec. \ref{sec:det-trjkt}. In this case, both detectors are accelerated in the same direction, i.e., $a_{1}$ and $a_{2}$ have the same sign. In this case, $q_{1}$ and $q_{2}$ also have the same signs. At the same time, we are going to assume that though the accelerations have the same signs they have different magnitudes, i.e., $a_{1}\neq a_{2}$. In this scenario, with the help of the geodesic interval from Eq. \eqref{eq:chordal-general} and with the Gaussian switching function $\chi(\tau) = e^{-{\tau^{2}}/{T^{2}}}$, the nonlocal term can be expressed as
    \begin{equation}
        \mathcal{I}_{\epsilon} = -\frac{i}{4\pi^{2}} \int^{\infty}_{-\infty}d \tau_{1}  \int_{-\infty}^{\infty}d\tau_{2} \, e^{-\frac{\tau_{1}^{2}}{T^{2}}} e^{-\frac{\tau_{2}^{2}}{T^{2}}} \frac{e^{i \omega (\tau_{1}+ \tau_{2})}}{2\left( \frac{1}{\Lambda} - \frac{\cosh(q_{1}\tau_{1}- q_{2}\tau_{2})}{q_{1}q_{2}} - \frac{1}{\Lambda}\frac{a_{1}a_{2}}{q_{1}q_{2}} \right) + i\epsilon }~.
    \end{equation}
    By transforming one of the Gaussian functions $e^{-{\tau_{1}^{2}}/{T^{2}}}$ in terms of its Fourier representation one can easily compute one of the integrals in the above expression. The details for carrying out this integration are provided in Appendix \ref{Appn:GaussSwtch-Ije}. The final expression for $\mathcal{I}_{\epsilon}$ should be obtained using numerical integration. It should also be noted that in the Minkowski background the expression of $\mathcal{I}_{\epsilon}$ can be very easily obtained by putting $\Lambda=0$ in the above expression. The explicit form for the Minkowski $\mathcal{I}_{\epsilon}$ is provided in Eq. \eqref{AppEq:IeM-Gauss-CI}.

%%%%%%%%%%%%%%%%%%%%%%%%%%%%%%%%%%%%%%%%%%%%%%%%%%%%%%%%%%%%%%%%%%%%%%%%%%%%%%%%%%%%%%%%%%%%%%%%%
\begin{figure}[!h]
\centering
\includegraphics[width=12.0cm]{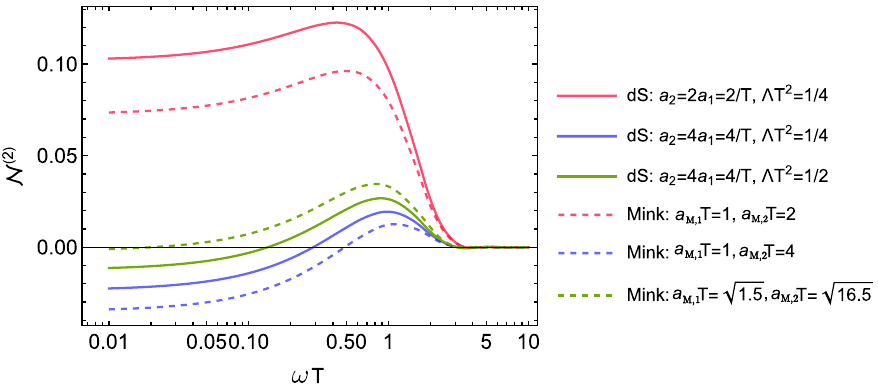}
\caption{\textbf{Gaussian switching, Case I:} In the above figure we provide the plot for Negativity for Case I, we plotted the Negativity for detectors in accelerated motion respectively in the de Sitter (Case I) and Minkowski backgrounds for Gaussian switching $\chi(\tau) = e^{-{\tau^{2}}/{T^{2}}}$. A comparison between the blue-thick and red-thick curves shows that entanglement decreases with the increase in acceleration keeping curvature $\Lambda$ fixed. Whereas, a comparison between the blue-thick and green-thick curves shows that entanglement increases with an increase in curvature $\Lambda$, while keeping the accelerations of the trajectories fixed. A similar conclusion holds for comparison between red-dashed and blue-dashed i.e. increasing acceleration decreases entanglement, hence \emph{degrading effect of increasing acceleration is independent of the presence of curvature}. The green-thick and green-dashed curves represent a comparison of the Negativity with $a_{M,j} = \sqrt{a_{j}^{2} + \Lambda}$ where $a_{j}$ is the dS acceleration. The comparison implies that entanglement in dS can not be mimicked by $a_{M,j} = \sqrt{a_{j}^{2} + \Lambda}$ and such a replacement actually overestimates the entanglement in dS spacetime. The comparison between (red, blue) curves for dS/Mink and green curves for dS/Mink, implies that $a_{M,j}=a_{j}$ underestimates the Negativity of dS in Minkowski spacetime, whereas $a_{M,j}=\sqrt{a_{M}^{2}+\Lambda}$ overestimates the Negativity in Minkowski spacetime.
}
\label{fig:NR-GaussSwitch-caseI}
\end{figure}

\begin{figure}[!h]
\centering
\includegraphics[width=12.0cm]{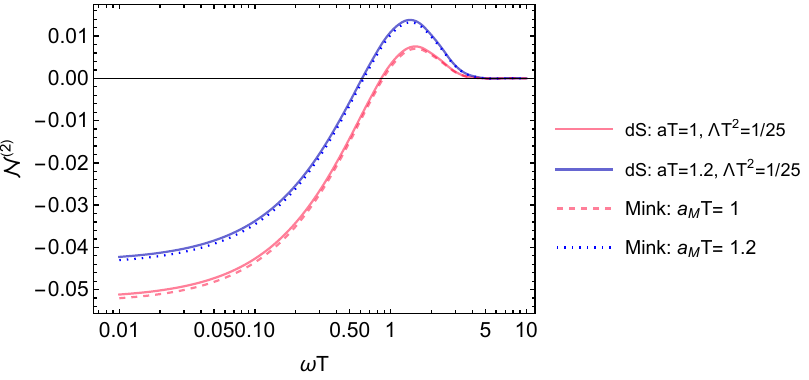}
\vskip 2 pt
\caption{\textbf{Gaussian switching, Case II:} In the above figure we provide the plot for Negativity mentioned in Case II of Sec. \ref{sec:det-trjkt} (where $q_1=-q_2=q$ and $a_1=-a_2=a$), we plotted the Negativity for detectors in accelerated motion respectively in the de Sitter (Case II) and Minkowski backgrounds for Gaussian switching $\chi(\tau) = e^{-{\tau^{2}}/{T^{2}}}$. It is evident from the plot that increasing acceleration has an enhancing effect in dS (blue-thick and red-thick) and Minkowski (blue-dashed and red-dashed) spacetime. 
} 
\label{fig:NR-GaussSwitch-caseII}
\end{figure}
%%%%%%%%%%%%%%%%%%%%%%%%%%%%%%%%%%%%%%%%%%%%%%%%%%%%%%%%%%%%%%%%%%%%%%%%%%%%%%%%%%%%%%%%%%%%%%%%%

\item \textbf{Case II:} This case corresponds to trajectories in different static patches with opposite 4 and 5 accelerations ($q_{1}=-q_{2},a_{1}=-a_{2}$). The simplified form of the nonlocal term  $\mathcal{I}_{\epsilon}$ in this scenario is 
\begin{equation}\label{eq:IeDS-Gauss-CII}
    \mathcal{I}_{\epsilon} =  -\frac{i}{4\pi^{2}} \int^{\infty}_{-\infty} d\tau_{1}\int^{\infty}_{-\infty}d\tau_{2} \, e^{-\frac{\tau_{1}^{2}}{T^{2}}} e^{-\frac{\tau_{2}^{2}}{T^{2}}} \frac{e^{i \omega (\tau_{1}+ \tau_{2})}}{ \frac{2}{a^{2}+\Lambda}\left( 1 + \cosh(\sqrt{a^{2}+\Lambda}(\tau_{1}+\tau_{2})) \right) + i\epsilon }~.
\end{equation}
This integral also can be evaluated in a similar spirit and finally gives a $k$ integral which needs numerical computation. In the above expression if one considers $\Lambda=0$, one can obtain the Minkowski nonlocal term corresponding to two detectors accelerated in opposite directions. One can look into Eq. \eqref{AppEq:IeM-Gauss-CII} for the Minkowski expression. It should also be noted that the Minkowski and the de Sitter nonlocal terms have the same functional forms in this scenario with Minkowski acceleration replaced by $\sqrt{a^2+\Lambda}$ in the de Sitter case.

\item \textbf{Case III:} In this case, the detector trajectories have accelerations such as $a_{1}=-a_{2}=a$ and $q_{1}=q_{2}=q$. Thus one can represent the nonlocal term as

\begin{equation}\label{eq:IeDS-Gauss-CIII}
      \mathcal{I}_{\epsilon} =  -\frac{i}{4\pi^{2}} \int^{\infty}_{-\infty} d\tau_{1}\int^{\infty}_{-\infty}d\tau_{2} \, e^{-\frac{\tau_{1}^{2}}{T^{2}}} e^{-\frac{\tau_{2}^{2}}{T^{2}}} \frac{e^{i \omega (\tau_{1}+ \tau_{2})}}{ \frac{2}{a^{2}+\Lambda}\left( \frac{2a^{2}+\Lambda}{\Lambda} + \cosh(\sqrt{a^{2}+\Lambda}(\tau_{1}-\tau_{2})) \right) + i\epsilon }~.
\end{equation}
In this scenario, by putting $\Lambda=0$ one cannot get the Minkowski nonlocal terms, as the trajectories do not have any correspondence to the Minkowski background. Here the best thing one consider is to take two detectors in opposite acceleration in the Minkowski background, for which the expression of the nonlocal term is provided in Eq. \eqref{AppEq:IeM-Gauss-CII}.

%%%%%%%%%%%%%%%%%%%%%%%%%%%%%%%%%%%%%%%%%%%%%%%%%%%%%%%%%%%%%%%%%%%%%%%%%%%%%%%%%%%%%%%%%%%%%%%%%
\begin{figure}[!h]\label{fig:34}
\centering
\includegraphics[width=12.0cm]{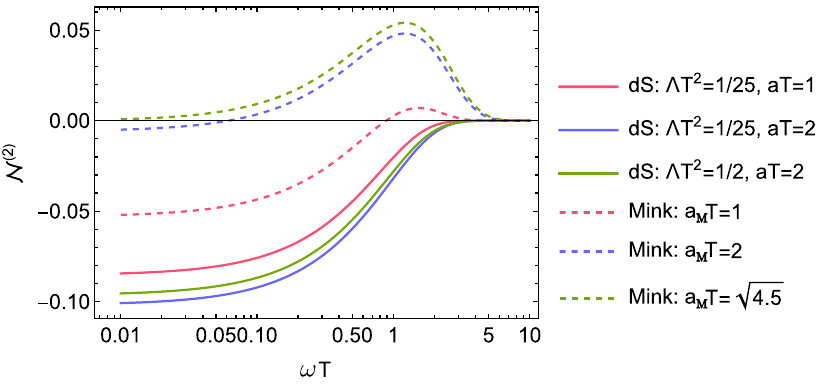}
\caption{\textbf{Gaussian switching, Case III:} In the above figure we calculated the Negativity for detector trajectories mentioned in Case III (where  $q_1=q_2=q$ and $a_1=-a_2=a$), we plotted the Negativity for detectors in accelerated motion respectively in the de Sitter (Case III) and Minkowski backgrounds for Gaussian switching $\chi(\tau) = e^{-{\tau^{2}}/{T^{2}}}$. A comparison between blue-thick and red-thick curves implies that increasing acceleration has a degrading effect on Negativity with curvature $\Lambda$ kept fixed. Whereas if we compare blue-thick and green-thick curves, it implies that increasing curvature has an enhancing effect on Negativity while keeping acceleration fixed. On the other hand in the Minkowski background, increasing acceleration has an enhancing effect on Negativity. On comparison of green-thick and green-dashed, we can infer that an attempt to mimic dS Negativity in Minkowski spacetime ($a_{M,j}=\sqrt{a_{j}^{2}+ \Lambda}$) overestimates the dS Negativity.  
} 
\label{fig:NR-GaussSwitch-caseIII}
\end{figure}

\begin{figure}[!h]
\centering
\includegraphics[width=12.0cm]{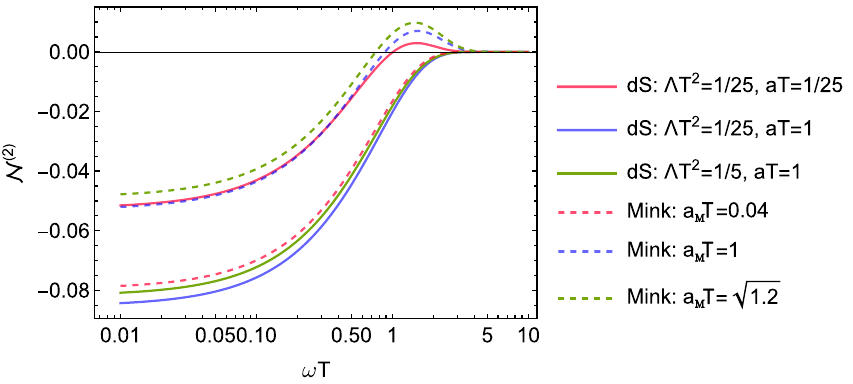}
\caption{\textbf{Gaussian switching, Case IV:} In the above figure we provide the plot for Negativity for detector trajectories mentioned in Case IV (where $q_1=-q_2=q$ and $a_1=a_2=a$), we plotted the Negativity for detectors in accelerated motion respectively in the de Sitter (Case IV) and Minkowski backgrounds for Gaussian switching $\chi(\tau) = e^{-{\tau^{2}}/{T^{2}}}$. On comparison of blue-thick and red-thick curves, we can infer that increasing acceleration has a degrading effect on entanglement keeping curvature $\Lambda$ fixed. Whereas, curvature has an enhancing effect on Negativity as can be seen from the comparison between blue-thick and green-thick curves. However, in Minkowski spacetime, increasing acceleration has an enhancing effect (blue-dashed and red-dashed). On comparison of green-thick and green-dashed, we can infer that an attempt to mimic dS Negativity in Minkowski spacetime ($a_{M,j}=\sqrt{a_{j}^{2}+ \Lambda}$) overestimates the dS Negativity.  
} 
\label{fig:NR-GaussSwitch-caseIV}
\end{figure}

\begin{figure}[h!]
\centering
\includegraphics[width=7.0cm]{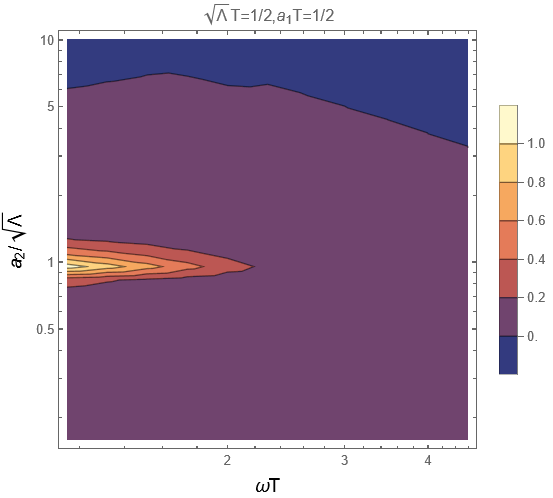}
\hskip 30pt
\includegraphics[width=7.0cm]{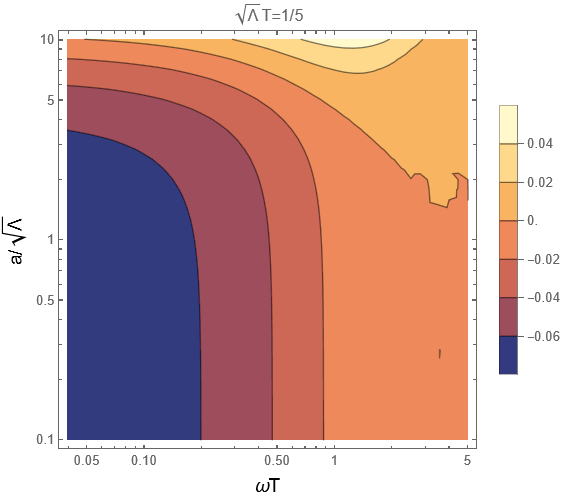}
\caption{In the above figure we present contour plots for Negativity rate $(\mathcal{N}^{(2)}/T)$ for a given parameter space.  \textbf{Left:} The contour plot corresponds to trajectories lying in the same static patch of \textbf{Case I}, where we have fixed the value of de-Sitter curvature $\sqrt{\Lambda}T=1/2$ and treated one of the detector trajectories as a reference trajectory $a_{1}T=1/2$. \textbf{Right:} The contour plot corresponds to \textbf{Case II}, where we have fixed the quantity $\sqrt{\Lambda}T$ $(=1/5)$ in such a way that the detectors are space-like separated.}
\label{fig:contour-CI-II}
\end{figure}

\begin{figure}[h!]
\centering
\includegraphics[width=7.0cm]{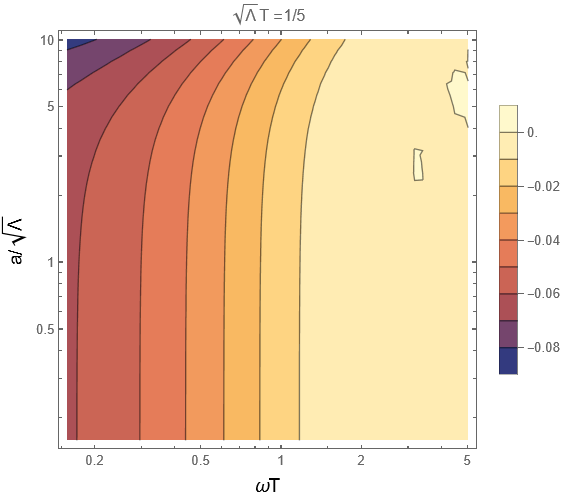}
\hskip 30pt
\includegraphics[width=7.0cm]{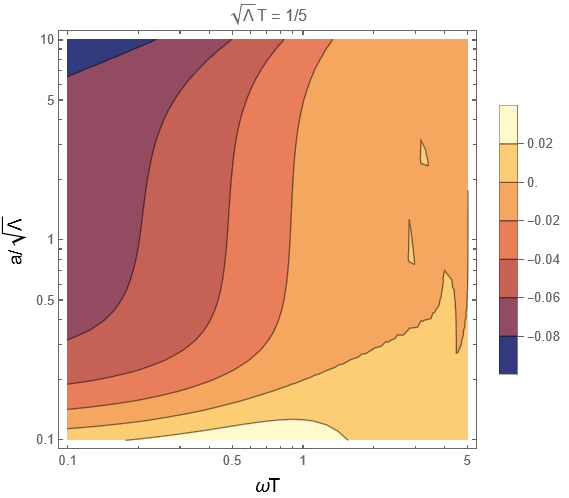}
\caption{The above figure depicts contour plots for Negativity rate $(\mathcal{N}^{(2)}/T)$ for given parameter spaces. \textbf{Left:} The contour plot corresponds to \textbf{Case III} (where $\sqrt{\Lambda}T=1/5$) with a large no-entanglement region except for high energy gaps $(\omega T)$ and acceleration $(a/\sqrt{\Lambda})$. \textbf{Right:} The contour plot corresponds to \textbf{Case IV} with $\sqrt{\Lambda}T=1/5$. The entanglement between the detectors can be seen for small acceleration ($a/\sqrt{\Lambda}$) and medium energy gaps ($\omega T$).}
\label{fig:contour-CIII-IV}
\end{figure}
%%%%%%%%%%%%%%%%%%%%%%%%%%%%%%%%%%%%%%%%%%%%%%%%%%%%%%%%%%%%%%%%%%%%%%%%%%%%%%%%%%%%%%%%%%%%%%%%%

\item \textbf{Case IV:} In this case trajectories are in different static patches with the same magnitude of acceleration but oriented in $\theta=0$ and $\theta = \pi$ planes respectively. The accelerations of the trajectories are related by, $a_{1}=a_{2}=a$ and $q_{1}=-q_{2}=q$. Thus the expression of the nonlocal term $\mathcal{I}_{\epsilon}$ with Gaussian switching can be written as
\begin{equation}\label{eq:IeDS-Gauss-CIV}
      \mathcal{I}_{\epsilon} =  -\frac{i}{4\pi^{2}} \int^{\infty}_{-\infty} d\tau_{1}\int^{\infty}_{-\infty}d\tau_{2}  e^{-\frac{\tau_{1}^{2}}{T^{2}}} e^{-\frac{\tau_{2}^{2}}{T^{2}}} \frac{e^{i \omega (\tau_{1}+ \tau_{2})}}{ \frac{2}{a^{2}+\Lambda}\left( \frac{2a^{2}+\Lambda}{\Lambda} - \cosh(\sqrt{a^{2}+\Lambda}(\tau_{1}-\tau_{2})) \right) + i\epsilon }~.
\end{equation}
Here also $\Lambda\to 0$ does not correspond to any physical scenario in the Minkowski background, i.e., here the transformation $a^{2}+\Lambda \to a^{2}$ does not produce the Minkowski $\mathcal{I}_{\epsilon}$ from the de Sitter one. The Minkowski nonlocal term should be obtained from Eq. \eqref{AppEq:IeM-Gauss-CII}, which corresponds to two detectors accelerated in opposite directions with the same magnitude of acceleration.

\end{itemize}

\subsubsection{Observations on Negativity}

In this part of the section, we provide analysis for Negativity corresponding to the four scenarios in de Sitter space as obtained in terms of $\mathcal{I}_{1}$ and $\mathcal{I}_{\epsilon}$ mentioned in the previous section. We also provide comparisons of these results with respect to the Minkowski Negativity. Here, we consider the particular situation when the detectors interact with the scalar field through a Gaussian window function $\chi(\tau) = e^{-\tau^{2}/T^{2}}$, and obtain the Negativity with numerical help utilizing the expression of \eqref{eq:neg}. The plots for Negativity corresponding to different cases are depicted in Figs. \ref{fig:NR-GaussSwitch-caseI}, \ref{fig:NR-GaussSwitch-caseII}, \ref{fig:NR-GaussSwitch-caseIII}, and \ref{fig:NR-GaussSwitch-caseIV}. Our key observations from these plots are as follows.
\begin{itemize}
    \item From Fig. \ref{fig:NR-GaussSwitch-caseI} corresponding to Case I of parallel accelerations, we observe that Negativity is non-vanishing, which is in contrast to the eternal switching scenario of Figs. \ref{fig:NR-eternal-caseII} and \ref{fig:NR-eternal-caseIV}. A comparison between the red-solid and the blue-solid curves shows that entanglement gets reduced with increasing acceleration of the second detector. Whereas, a comparison between the blue-solid and green-solid curves shows that Negativity is enhanced with increasing curvature. Comparison of a de Sitter curve with the Minkowski one (green-solid and green-dashed) asserts that Negativity in a de Sitter background is lesser compared to the Minkowski background, if the Minkowski accelerations are replaced by $\sqrt{\Lambda+a^2}$. Therefore, we conclude that in Case I, entanglement between the Minkowski and de Sitter spacetimes are not analogous concerning the mapping of $a$ with $\sqrt{\Lambda+a^2}$.
    
    \item For Case II, Fig. \ref{fig:NR-GaussSwitch-caseII} shows that with increasing acceleration and curvature the Negativity increases. Moreover, one can also observe that in this scenario the de Sitter Negativity is given exactly by the Minkowski result with the acceleration $(a)$ replaced by $\sqrt{a^2+\Lambda}$, which is actually guaranteed by the structure of the relevant Green's function. This implies even with the Gaussian switching one cannot obtain any observable effect in Case II as compared to the Minkowski background, and hence trajectories of Case II are not useful to probe the effects of curvature on entanglement. A new observation compared to the eternal switching is that entanglement in this case can vanish for small energy gaps. In fact, Negativity is maximum around the scale of $\omega\sim(1/T)$.
    
    \item For Case III, from Fig. \ref{fig:NR-GaussSwitch-caseIII} we observe that the de Sitter Negativity is lesser than the Minkowski Negativity with the Minkowski acceleration $\sqrt{a^2+\Lambda}$. With increasing acceleration of the detectors the Minkowski Negativity increases (red-dashed $\to$ blue-dashed $\to$ green-dashed curves), while the de Sitter Negativity decreases (red-solid $\to$ blue-solid curves). However, with increasing curvature the de Sitter Negativity gets enhanced, see the curves blue-solid and green-solid. We would like to point out that for our considered parameter values there is \textit{no entanglement} in the de Sitter scenario in this case.
    
    \item From Fig. \ref{fig:NR-GaussSwitch-caseIV} we observe that in Case IV the de Sitter entanglement can be greater or lesser compared to the Minkowski entanglement for Minkowski acceleration $\sqrt{a^2+\Lambda}$. For instance, for very small four-acceleration $(a)$, the de Sitter entanglement can be larger than the Minkowski one, see the red-solid and red-dashed curves. Whereas, for larger four-acceleration $(a)$ the de Sitter entanglement is always lesser than the Minkowski one, see blue-solid and blue-dashed curves. We find increasing acceleration reduces de Sitter entanglement (red-solid $\to$ blue-solid curves), while curvature has an enhancing effect on the entanglement (blue-solid $\to$ green-solid curves).

    \item Therefore, our observations for finite Gaussian switching also suggest that in a de Sitter background \textbf{\emph{presence of curvature (as quantified by curvature constant $\Lambda$) enhances entanglement for accelerated probes}}. On the other hand, \textbf{\emph{depending on specific configurations of acceleration, entanglement can be enhanced or reduced}}.

    \item In Cases I, III, and IV we observe that the entanglements in de Sitter are not analogous to the Minkowski entanglement concerning the replacement of the Minkowski four acceleration with $\sqrt{a^2+\Lambda}$. It should be noted that while Case I and Case III may correspond to a single cosmological patch in de Sitter, Cases II and IV do not denote the same. Therefore, to discern the effects of curvature in entanglement, Case I with finite switching seems to be the most useful scenario.

    \item In the Negativity contour plot for Case I, (see the left plot of Fig. \ref{fig:contour-CI-II}). We have considered one of the accelerated trajectories as the reference trajectory while varying the acceleration of the other trajectory. It can be inferred that entanglement is possible for values of acceleration in a small band around the reference trajectory which can be attributed mostly to causal communication.

    \item In the contour plot for Case II from Fig. \ref{fig:contour-CI-II}, we observed positive values for Negativity at large acceleration scales. However in Case IV from Fig. \ref{fig:contour-CIII-IV}, we observe that acceleration actually degrades entanglement.

    \item From Fig. \ref{fig:contour-CIII-IV}, we observe that no entanglement is possible in Case III for the considered parameter range. Moreover, in this scenario, for the chosen parameter values of the de Sitter length scale and acceleration, there seems to be no appreciable effect on entanglement.
    
    \item The Negativity in de Sitter spacetime can not be mimicked in Minkowski spacetime by replacing acceleration as $a_{M}= \sqrt{a^{2}+\Lambda}$ in Case I, III and IV, see Figs. \ref{fig:NR-GaussSwitch-caseI}, \ref{fig:NR-GaussSwitch-caseIII} and \ref{fig:NR-GaussSwitch-caseIV}. More precisely, such a replacement overestimates the Negativity compared to de Sitter spacetime. Whereas it can be mimicked in Case II as can be seen from the explicit expression of Negativity, see Eq. \eqref{AppEq:C2-Keq1-Ie-3}.

\end{itemize}

%%%%%%%%%%%%%%%%%%%%%%%%%%%%%%%%%%%%%%%%%%%%%%%%%%%%%%%%%%%5
\section{Analysis of the maxima in Negativity} \label{sec:Anlys-Maxima}

All of our above observations indicate that the entanglement for accelerated detectors in a de Sitter background has maxima concerning the energy gap. For an analysis of these maxima, we will consider the analytical expression of Negativity for eternal switching with the help of the expressions from Eqs. \eqref{eq:Keq1-Rj-1}, \eqref{AppEq:C2-Keq1-Ie-3} and \eqref{AppEq:C4-Keq1-Ie-2}. We will further notice that the maxima are obtained from a transcendental equation, which cannot be solved analytically. However, with numerical help, one can interpret the position of these maxima and understand the nature of entanglement at these parameter values. On the other hand, as Negativity itself is estimated numerically for finite Gaussian switching, it is very challenging to provide an analysis of the arriving peaks in this scenario. Let us go through the analysis of these peaks in the following paragraphs.

\subsection{Minkowski background}

We consider two detectors moving in opposite directions with uniform acceleration in Minkowski background. Thus, one of the detectors is confined in the right Rindler wedge, and the other one is in the left Rindler wedge. We also consider that they have accelerations of the same magnitude. This is analogous to Case II in de Sitter spacetime; see Fig. \ref{fig:dS-spacetime-embedded}. In the Minkowski spacetime, the expressions of the local and nonlocal terms in Negativity are obtained from Eqs. \eqref{eq:Keq1-Rj-1} and \eqref{AppEq:C2-Keq1-Ie-3} with $q$ replaced by the Minkowski $4-$acceleration $a$. Here, the expression of the Negativity rate is given by
\begin{eqnarray}\label{eq:NR-Mink-1}
    \mathcal{N}_{\mathcal{R}} &=& \frac{\omega}{2 \pi}\frac{ e^{\frac{\pi  \omega}{a}}-2}{\left(e^{\frac{2 \pi  \omega}{a}}-1\right)}~.
\end{eqnarray}
%
%%%%%%%%%%%%%%%%%%%%%%%%%%%%%%%%%%%%%%%%%%%%%%%%%%%%%%%%%%%%%%%%%%%%%%%%%%%%%%%%%%%%%%%%%%%%%%%%%
\begin{figure}[h!]
\centering
\includegraphics[width=8.0cm]{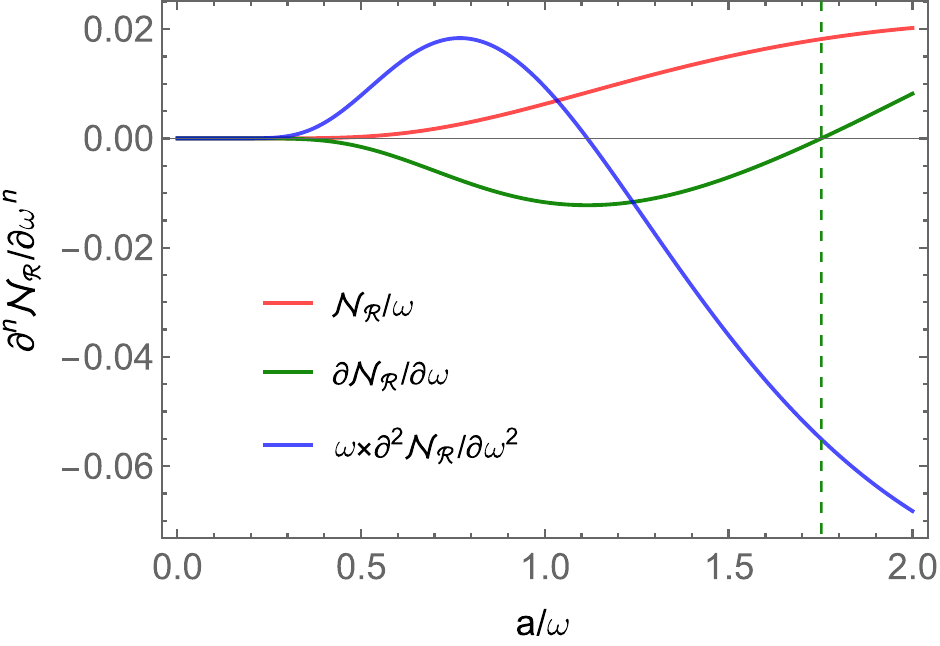}
\hskip 30pt
\includegraphics[width=8.0cm]{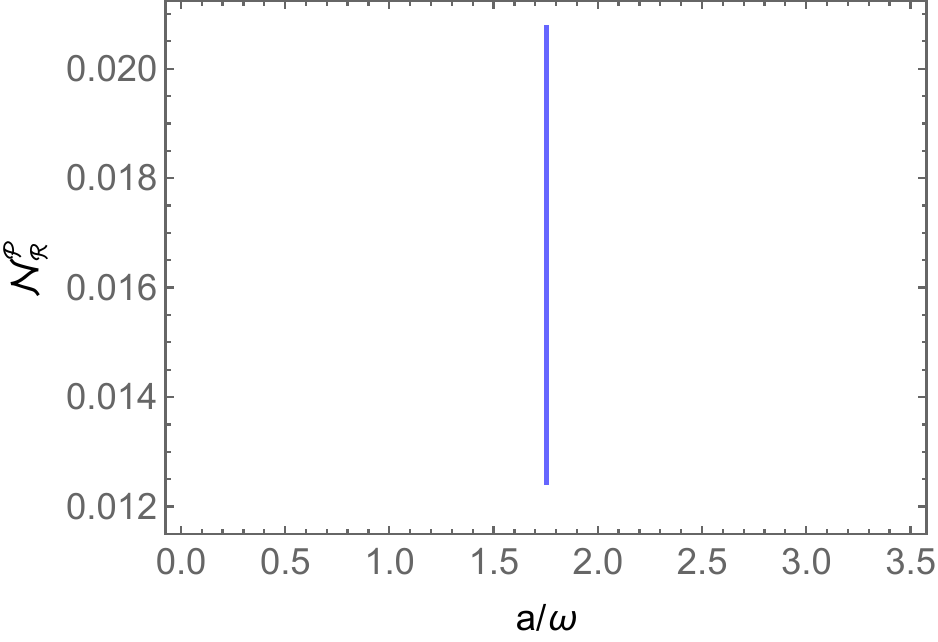}
\caption{{In the left we have plotted $\mathcal{N}_{\mathcal{R}}$ and its derivatives with respect to $\omega$ in Minkowski background as functions of the dimensionless parameter $\Bar{a}=a/\omega$. When $\partial\mathcal{N}_{\mathcal{R}}/\partial\omega$ vanishes, it corresponds to the maxima of the Negativity rate. The left part of the curve, when $\Bar{a}\lesssim 0.5$, does not correspond to any maxima as is evident from the plot of $\partial^2 \mathcal{N}_{\mathcal{R}} /\partial\omega^2$. On the right, we have plotted the dimensionless Negativity rate $\mathcal{N}^{\mathcal{P}}_{\mathcal{R}}$ that is evaluated at the peaks. Notice the peaks arrive at a fixed value of $\Bar{a}\sim 1.752$ (from left and right), which corresponds to a fixed value of $\mathcal{N}^{\mathcal{P}}_{\mathcal{R}}/\omega\sim0.018$. Here, we would also like to mention that the probes are in uniform acceleration in opposite directions, and this scenario is analogous to Case II from the de Sitter Negativity analysis.}}
\label{fig:peaks-Mink}
\end{figure}
%%%%%%%%%%%%%%%%%%%%%%%%%%%%%%%%%%%%%%%%%%%%%%%%%%%%%%%%%%%%%%%%%%%%%%%%%%%%%%%%%%%%%%%%%%%%%%%%%
We define a dimensionless parameter $\Bar{a}=a/\omega$, and in terms of this parameter the first derivative of the above Negativity with respect to the energy gap becomes
\begin{eqnarray}\label{eq:DNR-Mink-1}
    \frac{d\mathcal{N}_{\mathcal{R}}}{d\omega} &=& \frac{2 \Bar{a}+e^{\frac{3 \pi }{\Bar{a}}} (\Bar{a}-\pi )-e^{\pi /\Bar{a}} (\Bar{a}+\pi )-2 e^{\frac{2 \pi }{\Bar{a}}} (\Bar{a}-2 \pi )}{2 \pi  \Bar{a} \left(e^{\frac{2 \pi }{\Bar{a}}}-1\right)^2}~.
\end{eqnarray}
From the above expression, one can obtain the value of $\Bar{a}$ for which $d\mathcal{N}_{\mathcal{R}}/d\omega$ vanishes, and this value is given by $\Bar{a}\sim 1.752$. One can check that at the above value the double derivative of $\mathcal{N}_{\mathcal{R}}$ with respect to the energy gap becomes negative, thus indicating it to be a maximum for the Negativity rate. One can further find that the ratio $\mathcal{N}_{\mathcal{R}}/\omega$ is a function of the dimensionless parameter $\Bar{a}$ only, and thus this ratio in a Minkowski background remains constant at the peaks even if $\omega$ or $a$ is varied. To be specific that ratio at the peaks has a fixed value $\mathcal{N}^{\mathcal{P}}_{\mathcal{R}}/\omega\sim0.018$. In Fig. \ref{fig:peaks-Mink} we have plotted the $d\mathcal{N}_{\mathcal{R}}/d\omega$ to obtain the peak values and also plotted $\mathcal{N}^{\mathcal{P}}_{\mathcal{R}}/\omega$ at these peaks.

\subsection{de Sitter background}

In this part, we shall consider the de Sitter Negativity for accelerated detectors. In particular, we will consider the particular scenario of Case II for eternal switching. In this regard, we obtain the Negativity rate $\mathcal{N}_{\mathcal{R}}$ with the help of the local and nonlocal terms from Eqs. \eqref{eq:Keq1-Rj-1} and \eqref{AppEq:C2-Keq1-Ie-3}. This Negativity rate is given by
\begin{eqnarray}\label{eq:NR-DS-CII-1}
    \mathcal{N}_{\mathcal{R}} &=& \frac{\omega \left(e^{\frac{\pi  \omega}{q}}-2\right)}{2 \pi  \left(e^{\frac{2 \pi  \omega}{q}}-1\right)}~,
\end{eqnarray}
%
%%%%%%%%%%%%%%%%%%%%%%%%%%%%%%%%%%%%%%%%%%%%%%%%%%%%%%%%%%%%%%%%%%%%%%%%%%%%%%%%%%%%%%%%%%%%%%%%%
\begin{figure}[h!]
\centering
\includegraphics[width=8.0cm]{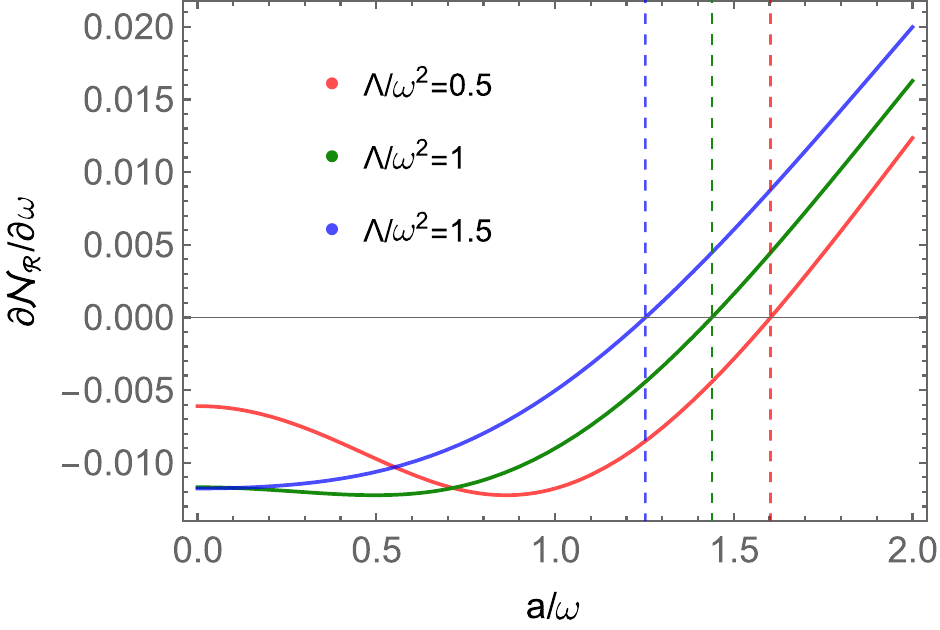}
\hskip 30pt
\includegraphics[width=8.0cm]{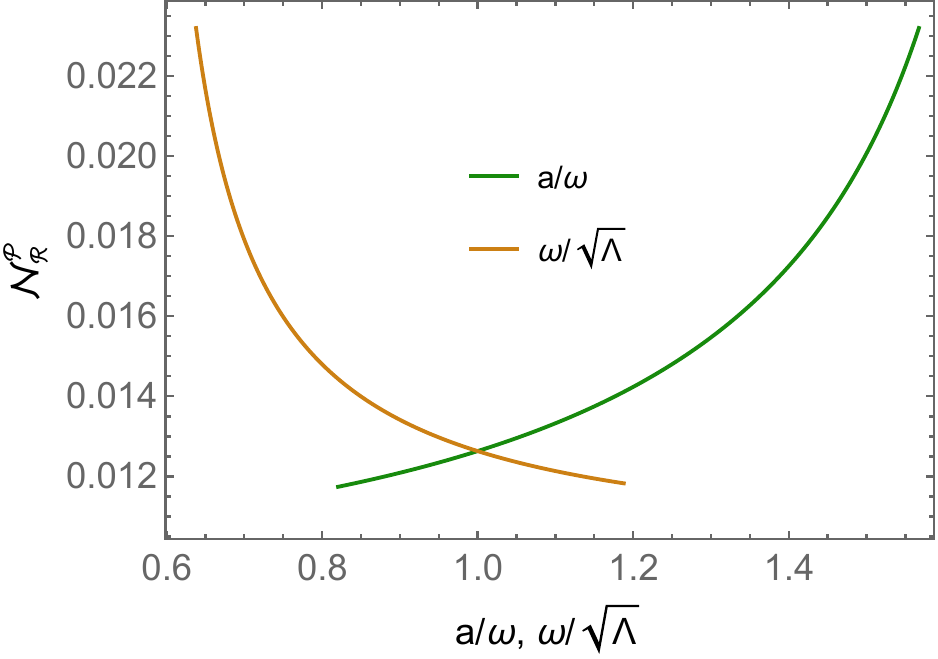}
\caption{On the left, we have plotted $\partial\mathcal{N}_{\mathcal{R}} /\partial\omega$ for de Sitter Case II as functions of the dimensionless parameters $\Bar{a}=a/\omega$. At the same time, on right, we have plotted the peak values of Negativity $\mathcal{N}^{\mathcal{P}}_{\mathcal{R}}$ as functions of the dimensionless variables $\Bar{a}=a/\omega$ and $\sqrt{\Bar{\Lambda}}=\sqrt{\Lambda}/\omega$. It is to be noted that here also $\mathcal{N}_{\mathcal{R}}/\omega$ is dimensionless and is described by the two dimensionless parameters $\Bar{a}$ and $\Bar{\Lambda}$. Unlike the Minkowski scenario, here the height of the peak Negativity, more precisely $\mathcal{N}^{\mathcal{P}}_{\mathcal{R}}/\omega$, increases as $\Bar{a}=a/\omega$ increases. On the other hand, we also observe that with increasing curvature, the $\mathcal{N}^{\mathcal{P}}_{\mathcal{R}}/\omega$ at the peaks decreases.}
\label{fig:peaks-DS-CII}
\end{figure}
\begin{figure}
    \centering
    \includegraphics[width=0.5\linewidth]{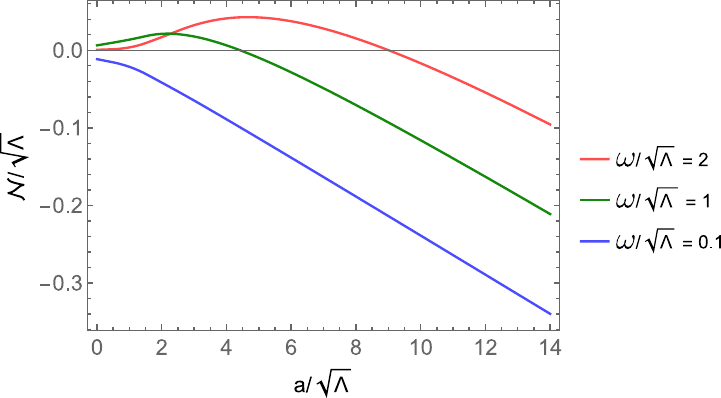}
    \caption{This figure indicates a peak in Negativity ($\mathcal{N}/\sqrt{\Lambda}$) for Case II of de Sitter as we vary the acceleration ($a/\sqrt{\Lambda}$), keeping $\omega/\sqrt{\Lambda}$ constant for each of the curves. As we decrease $\omega/\sqrt{\Lambda}$, the peak in Negativity disappears and it becomes negative for all values of $a/\sqrt{\Lambda}$.}
    \label{fig:dS-neg-acc-peak}
\end{figure}
%%%%%%%%%%%%%%%%%%%%%%%%%%%%%%%%%%%%%%%%%%%%%%%%%%%%%%%%%%%%%%%%%%%%%%%%%%%%%%%%%%%%%%%%%%%%%%%%%
where, $q=\sqrt{a^2+\Lambda}$. We will be considering only the Case II scenario as the Negativity, in this case, has a functional form similar to the Minkowski background, and the identification of a difference in terms of the peaks as the curvature is introduced is of interest to us. It should be noted that all other cases are much different from the Minkowski Negativity and thus may not require an explicit analysis of the peaks in the present context. We define the dimensionless parameters $\Bar{a}=a/\omega$ and $\Bar{\Lambda}=\Lambda/\omega^{2}$. Then the derivative of the above Negativity rate with respect to the energy gap yields
\begin{eqnarray}\label{eq:DNR-DS-CII-1}
    \frac{d\mathcal{N}_{\mathcal{R}}}{d\omega} &=& \frac{\left\{\sqrt{\Bar{a}^2+\Bar{\Lambda} }-\pi   \coth \left(\frac{\pi  }{\sqrt{\Bar{a}^2+\Bar{\Lambda} }}\right)\right\} \mathrm{csch}\left(\frac{\pi  }{\sqrt{\Bar{a}^2+\Bar{\Lambda} }}\right)}{4 \pi  \sqrt{\Bar{a}^2+\Bar{\Lambda} }}~.
\end{eqnarray}
The above equation denotes a transcendental equation in terms of $\Bar{a}$ and $\Bar{\Lambda}$, and thus should be solved numerically to obtain the roots. Compared to the previous Minkowski scenario one can understand that now these roots will also depend on the value of $\Bar{\Lambda}$  in addition to depending on $\Bar{a}$, see Fig. \ref{fig:peaks-DS-CII}. In Fig. \ref{fig:peaks-DS-CII} we have also plotted the ratio $\mathcal{N}^{\mathcal{P}}_{\mathcal{R}}/\omega$ at the peaks as functions of the dimensionless parameters $\Bar{a}$ and $\Bar{\Lambda}$. From this figure, we observe that the peak height, specifically $\mathcal{N}^{\mathcal{P}}_{\mathcal{R}}/\omega$, increases with increasing acceleration and curvature. 
The Negativity also has a peak with respect to the acceleration as plotted in Fig. \ref{fig:dS-neg-acc-peak}, i.e., as the acceleration is increased the Negativity first increases and then starts to decrease.

%%%%%%%%%%%%%%%%%%%%%%%%%%%%%%%%%%%%%%%%%%%%%%%%%%%%%%%%%%%%%%%%%%%%%%%%

\section{Discussion}\label{sec:concluding-sec}
%%%%%%%%%%%%%%%%%%%%%%%%%%%%%%%%%%%%%%%%%%%%%%%%%%%%%%%%%%%%%%%%%%%%%%%%

\noindent In this section, we briefly comment on various key aspects of the results.
\begin{enumerate}
    \item Let us start by considering the geodesic limit of our accelerated congruence, which requires some care. In this limit, we expect to recover the results already obtained in Ref. \cite{K:2023oon}. 
The initial separation of the accelerated trajectories in the small acceleration limit can be obtained by evaluating the geodesic distance $\sigma_{\text{geod}}$ from Eq. \eqref{eq:sgm-geod} by putting $\tau_{1}=\tau_{2}=0$ in the expression of Eq. \eqref{eq:chordal-general}. This initial separation is obtained to be 
\begin{eqnarray}\label{eq:congruence}
    d_0=\frac{2}{\sqrt{\Lambda}}\sin^{-1}\l[\frac{\sqrt{\Lambda}}{2}\sqrt{\l(\frac{2}{\Lambda}-\frac{2}{q_1\,q_2}\l(1+\frac{a_1\,a_2}{\Lambda}\r)\r)}\r]~.
\end{eqnarray}
It can be checked explicitly that in the limit of small acceleration and large $\Lambda$ values, the above expression reduces to the one in \cite{K:2023oon}. Consequently, the Negativity as obtained in Fig. \ref{fig:small-acc-limit} comes out to be similar to the ones obtained in Ref. \cite{K:2023oon}, without any bumps.

%%%%%%%%%%%%%%%%%%%%%%%%%%%%%%%%%%%%%%%%%%%%%%%%%%%%%%%%%%%%%%%%%%%%%%%%%%%%%%%%%%%%%%%%%%%%%%%%%
\begin{figure}[h!]
\centering
\includegraphics[width=8.0cm]{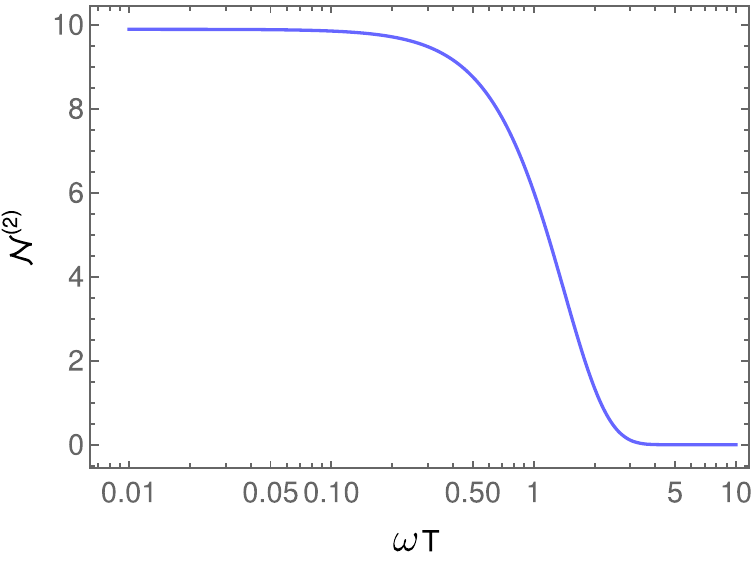}
\caption{The Negativity is plotted in the small acceleration limit with $\Lambda\,T^2=1$, $a_1=10^{-2} \,a_2 = 2\times10^{-2}$. One can observe that the bump disappears and the curve is similar to the one obtained in Ref. \cite{K:2023oon} for the case of geodesic detectors with initial separation $d_0/T\approx 10^{-2}$.} 
\label{fig:small-acc-limit}
\end{figure}
%%%%%%%%%%%%%%%%%%%%%%%%%%%%%%%%%%%%%%%%%%%%%%%%%%%%%%%%%%%%%%%%%%%%%%%%%%%%%%%%%%%%%%%%%%%%%%%%%

\item We would also like to point out another interesting fact about our accelerated congruence. We can rearrange Eq. \eqref{eq:congruence} to solve for $a_1=a$, for a given $a_2=a_{0}$ (considered as a reference curve). The corresponding expression turns out to be the same as the one obtained in \cite{Kothawala:2024dfk} by imposing \textit{rigidity condition} on congruences in curved spacetimes:
\begin{eqnarray}\label{eq:rigidity}
    a_{n} =  \sqrt{\Lambda} \tan \left(\sqrt{\Lambda}d_{n} - \tan^{-1}\frac{a_{0}}{\sqrt{\Lambda}}  \right)~.
\end{eqnarray}
The above expression yields the Rindler congruence in the limit $\Lambda \to 0$. This suggests an interesting connection between the rigidity condition employed in \cite{Kothawala:2024dfk} and uniformly accelerated congruences in a maximally symmetric spacetimes.
%%%%%%%%%%%%%%%%%%%%%%%%%%%%%%%%%%%%%%%%%%%%%%%%%%%%%%%%%%%%%%%%%%%%%%%%

\item Finally, we list and discuss some generic features of our observations. In particular, we observe that for eternal switching, detectors only in anti-parallel acceleration can have non-vanishing entanglement between them. Whereas, for finite Gaussian switching even detectors in parallel acceleration can get entangled. Our generic observations in both the eternal and finite switching scenarios are as follows.

\begin{enumerate}
    \item For single detector transition probabilities, see Eqs. \eqref{eq:Keq1-Rj-1}, the curvature and acceleration play similar roles. Single detector response is governed by the overall quantity $(a^2+\Lambda)^{1/2}$ and increases with increasing curvature and acceleration. We would like to mention that this is not an unknown phenomenon and it was previously observed in \cite{deser1997accelerated, hari:2021gns}.

    \item Unlike single detector transition probabilities, acceleration and curvature do not always affect entanglement in a similar manner. For instance, with our current set-up in a de Sitter background \textbf{\emph{curvature has an enhancing effect on entanglement.}} Whereas, \textbf{\emph{acceleration can enhance or inhibit entanglement}} depending on different scenarios that in turn depend on the relative direction of acceleration between the probes. It also indicates that substitution of the Minkowski acceleration with $(\Lambda+a^2)^{1/2}$ will not readily produce the de Sitter entanglement in every scenario. However, such an analogy between Minkowski and de Sitter spaces does exist for the de Sitter set-up of Case II. The analysis of Appendix \ref{Appn:negativity-same-wedge}, where we do not consider trajectories that obey a particular congruence of accelerated curves but two translated Rindler trajectories with the same acceleration, also corroborates these observations.

    \item Our observations also indicate that the entanglement for accelerated probes in a de Sitter background has maxima with respect to the energy gap of the detectors. We have seen using numerical methods that it is possible to understand some of the features of these Negativity peaks considering the analytical expressions of eternal switching, see Eqs. \eqref{eq:DNR-Mink-1} and \eqref{eq:DNR-DS-CII-1}. With a nontrivial switching function, the Negativity itself is estimated numerically, and it becomes a challenge to provide an analysis of the peaks present in this scenario. 

    \item For instance, the maxima of the Minkowski entanglement is decided by a single parameter $(a/\omega)$, and it is achieved at $(a/\omega)\sim 1.752$. In fact, one can observe that the ratio $\mathcal{N}^{\mathcal{P}}_{\mathcal{R}}/\omega$ remains constant at the peaks for different acceleration $a$ in the Minkowski background, specifically at $\mathcal{N}^{\mathcal{P}}_{\mathcal{R}}/\omega\sim 0.018$. On the other hand, for Case II of the de Sitter scenario, which happens to be functionally analogous to the Minkowski case regarding entanglement, the ratio $\mathcal{N}_{\mathcal{R}}/\omega$ has the same feature as Minkowski, but with respect to $q/\omega=\sqrt{a^2+\Lambda}/\omega$, which depends on the curvature length scale $\Lambda$. Similar features have been noted in the context of massive complex scalar fields in \cite{Perche:2021clp}. This highlights how curvature can affect quantum processes even when the curvature length scale is smaller than the acceleration length scale, which is even true in the case of a single detector (see Ref. \cite{hari:2021gns,K:2023dmj}). Such characteristics that are sensitive to curvature can help to develop a scheme to reconstruct spacetime from quantum measurements (for example, see Ref. \cite{K:2023oon, Kothawala:2023fkl, Padmanabhan:1996ap, Kempf:2021xlw}).
    
\end{enumerate}

\end{enumerate}

\noindent Our present work derives its significance from the fact that it enables us to distinguish between the effects of curvature and acceleration through entanglement measures. It has a significant scope to be implemented in generalized curved spacetime, see \cite{K:2023oon} for a similar endeavour with entanglement between geodesic probes. However, unlike the de Sitter case, in a generalized curved spacetime, curvature might inhibit entanglement.
From observational and experimental point of view, it might be possible to implement a similar set-up as ours in analogue systems, see \cite{Sanchez-Kuntz:2022gds, Barman:2022utm, Schmidt:2024zpg, Agullo:2024lry}. This will help validate some of the key results presented in this work.

\section*{Acknowledgements}
    Mayank and H.K. thank the Indian Institute of Technology (IIT) Madras and the Ministry of Human Resources and Development (MHRD), India for financial support. S.B. would like to thank the Science and Engineering Research Board (SERB), Government of India (GoI), for supporting this work through the National Post Doctoral Fellowship (N-PDF, File number: PDF/2022/000428).

%\newpage
\appendix

\section{Different integrals for eternal switching}\label{Appn:EternalSwtch-Ije}

In this section of the appendix, we explicitly evaluate the integrals $\mathcal{I}_{1}$ and $\mathcal{I}_{\varepsilon}$ for the eternal switching scenario. In this regard, we shall consider the expressions of the Wightman function and the Feynman propagators from Eq. \eqref{eq:gen-Gw-Gf}, and the expression of Eq. \eqref{eq:Ij-Ie-general} with $\chi(\tau)=1$. In particular, these expressions will help us evaluate the rates $\mathcal{R}_{1}$ and $\mathcal{R}_{\epsilon}$.

\subsection{Evaluation of the local terms in Negativity}

The Wightman function for detectors in accelerated trajectories in a de Sitter background is obtained from Eqs. \eqref{eq:chordal-general} and \eqref{eq:gen-Wightman}. In particular, the explicit expression of the Wightman function for the first detector is given by 
\begin{eqnarray}\label{AppEq:Ij-GreenFn}
    \mathcal{G}_{W}(\tau_{1},\tau'_{1}) = -\frac{1}{4\pi^{2}}\,\frac{q^2_{1}}{4 \sinh ^2\left\{\frac{q_{1} }{2}(\tau_{1}-\tau'_{1})-i \epsilon \right\}}~.
\end{eqnarray}
We consider a change of variables as $u=\tau_{1}-\tau'_{1}$ and $\tau=\tau_{1}$. Then from Eq. (\ref{eq:Ij-general}) one can find the local terms $\mathcal{I}_{1}$ as 
\begin{eqnarray}\label{AppEq:Keq1-Ij-1}
    \mathcal{I}_{1} &=& \int_{-\infty}^{\infty}d\tau\int_{-\infty}^{\infty}du\, e^{-i\,\omega\,u}\,\mathcal{G}_{W}(u,v)~\nonumber\\
    ~&=& -\frac{q^2_{1}}{16\pi^{2}}\,\int_{-\infty}^{\infty}d\tau\int_{-\infty}^{\infty}du\,\frac{e^{-i\,\omega\,u}}{\sinh ^2\left\{\frac{q_{1} \,u}{2}-i \epsilon \right\}}\nonumber\\
    ~&=& \frac{\omega}{\pi}\,\frac{1}{e^{2 \pi  \omega/q}-1}\,\int_{-\infty}^{\infty}d\tau~.
\end{eqnarray}
It is to be noted that the above quantity signifies the individual detector transition probability. We define the rate of this transition as
\begin{eqnarray}
    \mathcal{R}_{1} = \frac{\mathcal{I}_{1}}{\lim_{\mathbb{T}\to\infty}\int_{-\mathbb{T}}^{\mathbb{T}}d\tau} = \frac{\omega}{\pi}\,\frac{1}{e^{2 \pi  \omega/q}-1}~.
\end{eqnarray}
The rate of the local term for the second detector can be obtained in a similar manner.

\subsection{Evaluation of the non-local term in Negativity}

In this regard, we consider the integral representation of $\mathcal{I}_{\varepsilon}$ from Eq. \eqref{eq:Ie-general}, and different detector trajectories.

\begin{itemize}
    \item \textbf{Case I:} First, we consider a pair of detector trajectories as specified in Case I of Sec. \ref{sec:Ent-Acltd-probes}. In this scenario, one can notice that for eternal switching only poles for $(\tau_{1}+\tau_{2})$ in the upper half complex plane will contribute to non-vanishing $\mathcal{I}_{\varepsilon}$. The denominator of the integral is a function of $(q_{2}\tau_{2}-q_{1}\tau_{1})$, and one can express it as
\begin{eqnarray}\label{AppEq:C1-Keq1-Ie-1}
    \mathcal{I}_{\varepsilon} &=& \frac{i}{8\pi^{2}}\,\int_{-\infty}^{\infty}d\tau_{1}\int_{-\infty}^{\infty}d\tau_{2}\,\frac{e^{i\,\omega\,(\tau_{1}+\tau_{2})}}{\left\{ \frac{1}{\Lambda} - \frac{a_{1}a_{2}}{\Lambda q_{1}q_{2}} -\frac{ \cosh(q_{2}\tau_{2}-q_{1}\tau_{1})}{q_{1}q_{2}} \right\}+i\,\epsilon/2}~.
\end{eqnarray}
We consider a change of variables $v=\tau_{1}+\tau_{2}$ and $\tau=\tau_{1}$. The Jacobian for this transformation is unity. Then the above integral can be expressed as 
\begin{eqnarray}\label{AppEq:C1-Keq1-Ie-2}
    \mathcal{I}_{\varepsilon} &=& \frac{i}{8\pi^{2}}\,\int_{-\infty}^{\infty}d\tau\int_{-\infty}^{\infty}dv\,\frac{e^{i\,\omega\,v}}{\left\{ \frac{1}{\Lambda} - \frac{a_{1}a_{2}}{\Lambda q_{1}q_{2}} -\frac{ \cosh(q_{2}\,v-(q_{1}+q_{2})\tau)}{q_{1}q_{2}} \right\}+i\,\epsilon/2}~,
\end{eqnarray}
which has poles at $v=\l[\pm\cosh ^{-1}\left\{(-a_{1} a_{2}+i \Lambda  q_{1} q_{2} \epsilon +q_{1} q_{2})/\Lambda \right\}+q_{1} \tau +q_{2} \tau +2\,i\,\pi\,n\r]/q_{2}$, where $n\in \mathbb{Z}$ with $\mathbb{Z}$ being the set of all integers. After carrying out the integrations and the sum over $n$ one gets the result 
\begin{eqnarray}\label{AppEq:C1-Keq1-Ie-3}
    \mathcal{I}_{\varepsilon} &=& \frac{\Lambda  q_{1} \left\{\coth \left(\frac{\pi  \omega}{q_{2}}\right)-1\right\} e^{\frac{\omega \,\pi}{q_{2}}} \sinh \left[\frac{\omega \left(\pi -i \cosh ^{-1}\left(\frac{q_{1} q_{2}-a_{1} a_{2}}{\Lambda }\right)\right)}{q_{2}}\right]}{4 \pi  \sqrt{(a_{1} a_{2}+\Lambda -q_{1} q_{2})(a_{1} a_{2}-\Lambda -q_{1} q_{2})}}\,\int_{-\infty}^{\infty}d\tau\,e^{\frac{ i \, \omega\,\tau  (q_{1}+q_{2})}{q_{2}}}~,
\end{eqnarray}
The integration over $\tau$ will provide a Dirac delta distribution $\delta( \omega\, (q_{1}+q_{2})/q_{2})$, which vanishes as both $q_{1}$ and $q_{2}$ are with the same sign for Case I.
\vspace{0.2cm}

\item \textbf{Case II:} Second, we consider the scenario when the two observers are on different static patches with the same magnitude of accelerations, though in opposite directions. The geodesic distance is obtained from Eq. (\ref{eq:C2-geodesic-dist}), and one can express the nonlocal term as 
\begin{eqnarray}\label{AppEq:C2-Keq1-Ie-1}
    \mathcal{I}_{\varepsilon} &=& \frac{i\,q^2}{16\pi^{2}}\,\int_{-\infty}^{\infty}d\tau_{1}\int_{-\infty}^{\infty}d\tau_{2}\,\frac{e^{i\,\omega\,(\tau_{1}+\tau_{2})}}{\cosh^2\l\{\frac{q}{2}(\tau_2+\tau_1)\r\}+i\,\epsilon/4}~.
\end{eqnarray}
We consider a change of variables to $v=\tau_{1}+\tau_{2}$ and $\tau=\tau_{1}$. The Jacobian for this transformation is $1$. Then from the previous equation, one can define the rate as 
\begin{eqnarray}\label{AppEq:C2-Keq1-Ie-2}
    \mathcal{R}_{\varepsilon} &=& \frac{i\,q^2}{16\pi^{2}}\,\int_{-\infty}^{\infty}dv\,\frac{e^{i\,\omega\,v}}{\cosh^2\l\{q\,v/2\r\}+i\,\epsilon/4}~.
\end{eqnarray}
One can find the poles for the integrand at $v= \l[2\,i\,\pi\,n\pm \cosh^{-1}\left\{\left(-2-i q^2 \epsilon \right)/2\right\}\r]/q$, where $n\in \mathbb{Z}$ with $\mathbb{Z}$ being the set of all integers. This information about the pole structure can now be utilized with the Residue theorem to obtain the expression of $\mathcal{R}_{\varepsilon}$. In particular, after the usage of the Residue theorem and then performing the sum over the integers $n$ one can obtain the final form of the previous expression as
\begin{eqnarray}\label{AppEq:C2-Keq1-Ie-3}
    \mathcal{R}_{\varepsilon} &=& \frac{i \, \omega \, e^{\pi  \omega/q}}{2 \, \pi \, \left(e^{2 \pi  \omega/q}-1\right)}~.
\end{eqnarray}
\vspace{0.2cm}

\item \textbf{Case III:} Third, we consider the scenario when the magnitudes of four and five accelerations for both the probes are the same in the two trajectories, but the four accelerations are in opposite directions. The expression of the geodesic interval for this scenario is provided in Eq. (\ref{eq:C3-geodesic-dist}). We use this expression to represent the nonlocal entangling terms as
\begin{eqnarray}\label{AppEq:C3-Keq1-Ie-1}
    \mathcal{I}_{\varepsilon} &=& \frac{i\,\Lambda\,q^2}{8\pi^{2}}\,\int_{-\infty}^{\infty}d\tau_{1}\int_{-\infty}^{\infty}d\tau_{2}\,\frac{e^{i\,\omega\,(\tau_{1}+\tau_{2})}}{\l\{q^2+a^2-\Lambda\,\cosh\l(q(\tau_2-\tau_1)\r)\r\}+i\,\epsilon\,\Lambda\,q^2/2}~.
\end{eqnarray}
With a change of variables to $v=\tau_{1}+\tau_{2}$ and $u=\tau_{2}-\tau_{1}$ one can easily check that $\mathcal{I}_{\varepsilon}$ vanishes. The reasoning behind this is simple: in the numerator of the integral, the exponential factor depends only on $v$, while the denominator depends only on $u$. Therefore, there is no pole for $v$ from the denominator, and thus, the numerator provides us with a Dirac delta distribution of $\delta(\omega)$, which vanishes for non-zero detector energy gap, i.e., for $\omega>0$.

Let us now prove the same using the change of variables $v=\tau_{1}+\tau_{2}$ and $\tau=\tau_{1}$. In particular, with this change of variables the integral from Eq. (\ref{AppEq:C3-Keq1-Ie-1}) looks like 
\begin{eqnarray}\label{AppEq:C3-Keq1-Ie-2}
    \mathcal{I}_{\varepsilon} &=& \frac{i\,\Lambda\,q^2}{8\pi^{2}}\,\int_{-\infty}^{\infty}d\tau\int_{-\infty}^{\infty}dv\,\frac{e^{i\,\omega\,v}}{2\,a^2+q^2 \,(2+i\,\Lambda  \,\epsilon )-2 \Lambda\, \cosh \{q\,(v-2 \tau )\}}~.
\end{eqnarray}
The poles are at $v=\l[\pm\cosh ^{-1}\left\{(2 \left(a^2+q^2\right)+i \Lambda  q^2 \epsilon)/(2 \Lambda)\right\}+2 q \tau +2\,i\,\pi\,n\r]/q$, where $n\in \mathbb{Z}$. Utilizing the Residue theorem and the sum over the integer $n$ we obtain
\begin{eqnarray}\label{AppEq:C3-Keq1-Ie-3}
    \mathcal{I}_{\varepsilon} &=& -\frac{\Lambda\,q  \,\text{csch}\left(\frac{\pi\,\omega}{q}\right) \sinh \left[\omega \left\{\pi -i \cosh ^{-1}\left(\frac{a^2+q^2}{\Lambda }\right)\right\}/q\right]}{8\,\pi  \sqrt{(a^2-\Lambda +q^2)\left(a^2+\Lambda +q^2\right) }}\,\int_{-\infty}^{\infty}d\tau\,e^{2\,i\,\tau\,\omega}~.
\end{eqnarray}
The integral over $\tau$ gives the Dirac delta distribution $\delta(2\,\omega)$, which makes the entire integral vanish as $\omega>0$.\vspace{0.2cm}

\item \textbf{Case IV:} Fourth, we consider a situation where the two detectors follow trajectories of the same four accelerations and the same but opposite five accelerations. The corresponding geodesic interval is obtained from Eq. (\ref{eq:C4-geodesic-dist}), and one can express the nonlocal entangling term as 
\begin{eqnarray}\label{AppEq:C4-Keq1-Ie-1}
    \mathcal{I}_{\varepsilon} &=& \frac{i}{4\pi^{2}}\,\int_{-\infty}^{\infty}d\tau_{1}\int_{-\infty}^{\infty}d\tau_{2}\,\frac{e^{i\,\omega\,(\tau_{1}+\tau_{2})}}{\frac{2}{q^2} \left\{\frac{a^2+q^2}{\Lambda }+\cosh (q v)\right\}+i \epsilon}~.
\end{eqnarray}
We consider a change of variables $v=\tau_{1}+\tau_{2}$ and $\tau=\tau_{1}$. Then poles are at $v= \left[2 i \pi  n\pm\cosh ^{-1}\{(-2 \left(a^2+q^2\right)-i \Lambda  q^2 \epsilon)/(2\,\Lambda) \}\right]/q$ for all $n\in \mathbb{Z}$. After using the Residue theorem and carrying out the sum over $n$ one can get the expression of $\mathcal{I}_{\varepsilon}$. In particular the rate $\mathcal{R}_{\varepsilon}$ is given by
\begin{eqnarray}\label{AppEq:C4-Keq1-Ie-2}
    \mathcal{R}_{\varepsilon} &=& \frac{\Lambda\,q\,e^{-\frac{i \omega \cosh ^{-1}\left(-\frac{a^2+q^2}{\Lambda }\right)}{q}} \left[-1+e^{2 \omega \left\{\pi +i \cosh ^{-1}\left(-\frac{a^2+q^2}{\Lambda }\right)\right\}/q}\right]}{4 \pi  \left(e^{\frac{2 \pi  \omega}{q}}-1\right) \sqrt{\left(a^2-\Lambda +q^2\right)\left(a^2+\Lambda +q^2\right)}}~.
\end{eqnarray}
\end{itemize}

\section{Different integrals for the Gaussian switching}\label{Appn:GaussSwtch-Ije}
In this section of the appendix, we provide some explicit expressions for the integrals $\mathcal{I}_{1}$ and $\mathcal{I}_{\varepsilon}$ for the Gaussian switching scenario. In this regard, we shall consider the expressions of the Wightman function and the Feynman propagators from Eq. \eqref{eq:gen-Gw-Gf}, and the expression of Eq. \eqref{eq:Ij-Ie-general} with $\chi(\tau)=e^{-{\tau^{2}}/{T^{2}}}$.

\subsection{Evaluation of the local terms in Negativity}

Here we evaluate the local terms $\mathcal{I}_{j}$ as specified in \eqref{eq:Ij-general} with the Gaussian switching. We consider a change of variables to $u=\tau_{j}-\tau'_{j}$ and $\tau=\tau_{j}$, and proceed to evaluate $\mathcal{I}_{j}$ in the Minkowski background. In particular, in the Minkowski spacetime, we have the following expression for these local terms,
\begin{equation}\label{AppEq:KeqG-IjM-1}
    \mathcal{I}_{j,M} = -\frac{1}{4 \pi^{2}}\int_{-\infty}^{\infty} dv\int_{-\infty}^{\infty}\, du \frac{e^{-i\omega u}}{\frac{4}{a^{2}}\sinh^2\l( \frac{a(u-i\epsilon)}{2} \r)} e^{-\frac{(v-u)^{2}}{T^{2}}}e^{-\frac{v^{2}}{T^{2}}} ~.
\end{equation}
It is evident from the above expression that the $v$ integral can be carried out easily and the remaining part is an integral over a Gaussian function of $u$. To perform this second integral over $u$ integral, we make use of Fourier decomposition of Gaussian functions, which is $e^{-u^{2}/T^{2}} = (T/\sqrt{2 \pi}) \int^{\infty}_{-\infty} dk \,e^{i k u - k^{2}T^{2}/2}$.
The utilization of this expression makes the exponential in our integral linear in $u$, and the $u$ integral can be easily done by appropriately choosing the contour and summing their respective residues. In the end, we are left with a $k$ integral which can be computed numerically using Mathematica \cite{Mathematica} with small $\epsilon$ values.
\begin{equation}\label{AppEq:KeqG-IjM-2}
    \mathcal{I}_{j,M} = \frac{T^{2}}{4 \pi}\int_{-\infty}^{\infty} dk \frac{e^{\frac{2k\pi}{a} - \frac{k^{2}T^{2}}{2} + (\omega - k)\epsilon}}{e^{\frac{2 k \pi}{a}} - e^{\frac{2 \pi \omega}{a}}}(k - \omega)
\end{equation}
The individual detector transition expression in de Sitter spacetime goes in complete parallel with the Minkowski case because of the similar structure of Wightman functions with the replacement of $a^{2} \longrightarrow a^{2} + \Lambda$. Hence we obtained the following expression in de Sitter spacetime,

\begin{equation}\label{AppEq:KeqG-IjDS-1}
    \mathcal{I}_{j,dS} = \frac{T^{2}}{4 \pi}\int_{-\infty}^{\infty} dk \frac{e^{\frac{2k\pi}{\sqrt{a^{2}+ \Lambda}} - \frac{k^{2}T^{2}}{2} + (\omega - k)\epsilon}}{e^{\frac{2 k \pi}{\sqrt{a^{2}+\Lambda}}} - e^{\frac{2 \pi \omega}{\sqrt{a^{2}+\Lambda}}}}(k - \omega)~.
\end{equation}

\subsection{Evaluation of the non-local term in Negativity}

There are four cases for non-local terms (terms denoting inter-detector correlation) that depend on the choice of accelerations for detector trajectories as discussed in Sec. \ref{sec:det-trjkt} and illustrated in fig \ref{fig:dS-spacetime-embedded}. Here, we estimate each case separately.
\begin{itemize}
    \item \textbf{Case I:} This case corresponds to two trajectories in the same static patch of the de Sitter spacetime with different magnitudes of acceleration. The $\mathcal{I}_{\epsilon,dS}$ integral to be evaluated simplifies to,

    \begin{equation}\label{AppEq:IeDS-Gauss-CI}
        \mathcal{I}_{\epsilon,dS} = \frac{-i}{4\pi^{2}} \int^{\infty}_{-\infty} d \tau_{1} d\tau_{2} \,  e^{-\frac{\tau_{1}^{2}}{T^{2}}} e^{-\frac{\tau_{2}^{2}}{T^{2}}} \frac{e^{i \omega (\tau_{1}+ \tau_{2})}}{2\left( \frac{1}{\Lambda} - \frac{\cosh(q_{1}\tau_{1}- q_{2}\tau_{2})}{q_{1}q_{2}} - \frac{1}{\Lambda}\frac{a_{1}a_{2}}{q_{1}q_{2}} \right) + i\epsilon }
    \end{equation}
We proceed in a similar way as done for $\mathcal{I}_{j}$ by Fourier decomposing the Gaussian function, and we get
\begin{equation}
    e^{-\frac{\tau_{1}^{2}}{T^{2}}} = \frac{T}{\sqrt{2 \pi}} \int^{\infty}_{-\infty} dk \,e^{i k \tau_{1} - \frac{k^{2}T^{2}}{2}} 
\end{equation}
and evaluating the $\tau_{1}$ integral by residues. The $\tau_{2}$ integral turns out to be a Gaussian integral which can be computed exactly and leaves us with a $k$ integral which can be computed numerically.\\
The corresponding expression in Minkowski spacetime evaluates inter-detector transitions for Rindler trajectories with different accelerations in the same Rindler patch.

\begin{equation}\label{AppEq:IeM-Gauss-CI}
    \mathcal{I}_{\epsilon,M} =   \frac{-i}{4\pi^{2}} \int^{\infty}_{-\infty} d\tau_{1}d\tau_{2} \,  e^{-\frac{\tau_{1}^{2}}{T^{2}}} e^{-\frac{\tau_{2}^{2}}{T^{2}}} \frac{e^{i \omega (\tau_{1}+ \tau_{2})}}{ \frac{1}{a_{1}^{2}} + \frac{1}{a_{2}^{2}} - \frac{\cosh(a_{1}\tau_{1}- a_{2}\tau_{2})}{a_{1}a_{2}} + i\epsilon }
\end{equation}

The numerical computation can be done in exact parallel, as elaborated for the de Sitter case. Note, however, that the structure of the Feynman propagator is different in both spacetimes, which results in a slight variation in Negativity. This also tells us that entanglement harvested in the de Sitter case can not be mimicked in Minkowski just by adjusting the 4-acceleration. In other words, it's a genuine curvature effect.

%%%%%%%%%%%%%%%%%%%%%%%%%%%%%%%%%%%%%%%%%%%%%%%%%%%%%%%%%%%%%%%%%%%%%%%%%%%%%%%%%%%%%%%%%%%%%%%%%

\item \textbf{Case II:} This case corresponds to trajectories in different static patches with opposite 4 and 5 accelerations ($q_{1}=-q_{2},a_{1}=-a_{2}$). The $\mathcal{I}_{\epsilon,dS}$ and $\mathcal{I}_{\epsilon,M}$ integrals simplifies to,
\begin{eqnarray}\label{AppEq:IeM-Gauss-CII}
    \mathcal{I}_{\epsilon,dS} &=&  \frac{-i}{4\pi^{2}} \int^{\infty}_{-\infty} \, d\tau_{1}d\tau_{2} e^{-\frac{\tau_{1}^{2}}{T^{2}}} e^{-\frac{\tau_{2}^{2}}{T^{2}}} \frac{e^{i \omega (\tau_{1}+ \tau_{2})}}{ \frac{2}{a^{2}+\Lambda}\left( 1 + \cosh(\sqrt{a^{2}+\Lambda}(\tau_{1}+\tau_{2})) \right) + i\epsilon }~\\
    \mathcal{I}_{\epsilon,M} &=&  \frac{-i}{4\pi^{2}} \int^{\infty}_{-\infty} d\tau_{1} d\tau_{2}  \, e^{-\frac{\tau_{1}^{2}}{T^{2}}} e^{-\frac{\tau_{2}^{2}}{T^{2}}} \frac{e^{i \omega (\tau_{1}+ \tau_{2})}}{ \frac{2}{a^{2}}\left( 1 + \cosh(a(\tau_{1}+\tau_{2})) \right) + i\epsilon }~.
\end{eqnarray}
Both of these integrals can be evaluated in a similar spirit and finally give a $k$ integral, which needs numerical computation. Note that in this case, both integrals share the same pole structure, and hence in this case, de Sitter's result can be mimicked by Minkowski trajectories just by interchanging the acceleration $a^{2}$ with $ a^{2}+ \Lambda$.

\item \textbf{Case III:} In this case, trajectories belong to the same static patch with the same magnitude of acceleration but lie in $\theta=0$ and $\theta = \pi$ planes, respectively. The corresponding accelerations are related by, $a_{1}=-a_{2}=a$ and $q_{1}=q_{2}=q$. Thus the expressions of $\mathcal{I}_{\epsilon,dS}$ and $\mathcal{I}_{\epsilon,M}$ can be simplified to,
\begin{eqnarray}
      \mathcal{I}_{\epsilon,dS} &=&  \frac{-i}{4\pi^{2}} \int^{\infty}_{-\infty} d\tau_{1}d\tau_{2} \, e^{-\frac{\tau_{1}^{2}}{T^{2}}} e^{-\frac{\tau_{2}^{2}}{T^{2}}} \frac{e^{i \omega (\tau_{1}+ \tau_{2})}}{ \frac{2}{a^{2}+\Lambda}\left( \frac{2a^{2}+\Lambda}{\Lambda} + \cosh(\sqrt{a^{2}+\Lambda}(\tau_{1}-\tau_{2})) \right) + i\epsilon }\\
    \mathcal{I}_{\epsilon,M} &=&  \frac{-i}{4\pi^{2}} \int^{\infty}_{-\infty} d\tau_{1}d\tau_{2} \, e^{-\frac{\tau_{1}^{2}}{T^{2}}} e^{-\frac{\tau_{2}^{2}}{T^{2}}} \frac{e^{i \omega (\tau_{1}+ \tau_{2})}}{ \frac{2}{a^{2}}\left( 1 + \cosh(a(\tau_{1}+\tau_{2})) \right) + i\epsilon }~.
\end{eqnarray}
Note here that a similar replacement in $\mathcal{I}_{\epsilon,M}$ of $a^{2} \to a^{2}+\Lambda$ does not reproduce de Sitter integral and hence trajectories in de Sitter case do probe a genuine effects of curvature on entanglement.

\item \textbf{Case IV:} In this case, trajectories are in different static patches with the same magnitude of acceleration but oriented in $\theta=0$ and $\theta = \pi$ planes, respectively. The accelerations on the trajectories are related as, $a_{1}=a_{2}=a$ and $q_{1}=-q_{2}=q$. Thus $\mathcal{I}_{\epsilon,dS}$ and $\mathcal{I}_{\epsilon,M}$ simplifies to,

\begin{eqnarray}
      \mathcal{I}_{\epsilon,dS} &=&  \frac{-i}{4\pi^{2}} \int^{\infty}_{-\infty} d\tau_{1}d\tau_{2} \,  e^{-\frac{\tau_{1}^{2}}{T^{2}}} e^{-\frac{\tau_{2}^{2}}{T^{2}}} \frac{e^{i \omega (\tau_{1}+ \tau_{2})}}{ \frac{2}{a^{2}+\Lambda}\left( \frac{2a^{2}+\Lambda}{\Lambda} - \cosh(\sqrt{a^{2}+\Lambda}(\tau_{1}-\tau_{2})) \right) + i\epsilon }\\
    \mathcal{I}_{\epsilon,M} &=&  \frac{-i}{4\pi^{2}} \int^{\infty}_{-\infty} d\tau_{1}d\tau_{2} \, e^{-\frac{\tau_{1}^{2}}{T^{2}}} e^{-\frac{\tau_{2}^{2}}{T^{2}}} \frac{e^{i \omega (\tau_{1}+ \tau_{2})}}{ \frac{2}{a^{2}}\left( 1 + \cosh(a(\tau_{1}+\tau_{2})) \right) + i\epsilon }~.
\end{eqnarray}

Note again that $a^{2} \to a^{2}+\Lambda$ in $\mathcal{I}_{\epsilon,M}$ does not reproduce $\mathcal{I}_{\epsilon,dS}$ starting from the Minkowski scenario.
\end{itemize}

\section{Negativity for accelerated trajectories in the same wedge}\label{Appn:negativity-same-wedge}

In the analysis considered so far, we only considered trajectories that obeyed a particular congruence of curves (for instance: Rindler congruence in Minkowski spacetime) and hence there is a direct correspondence between acceleration and distance dictated by the congruence. This restricts the analysis by the fact that one cannot consider 2 accelerated curves with the same acceleration separated by an initial distance $d$. Hence in this section, we consider this particular scenario and carry out the analysis of Negativity and also verify the specific limits. 

 \subsection{Minkowski background}
\begin{figure}[h!]
\centering
\includegraphics[width=12.0cm]{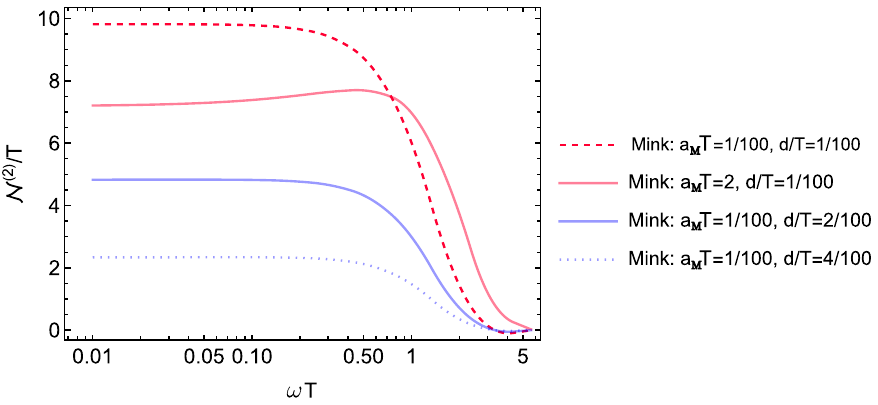}
\caption{The Negativity rate $(\mathcal{N}^{(2)}/T)$ is plotted against $\omega T$ between 2 trajectories with same acceleration $aT$ separated by an initial distance $d$ at $\tau_{1}=\tau_{2}=0$ in  Minkowski spacetime. Red curves show an increase in Negativity for a given initial separation ($d/T=10^{-2}$) and Negativity approaches a stationary detector limit. Blue curves show an increase in Negativity with decreasing initial separation for uniform acceleration $aT=10^{-2}$.  }
\label{fig:non-rindler-mink}
\end{figure}

First, we consider Negativity for accelerated trajectories having the same uniform acceleration $a$ in Minkowski spacetime. We solved the equation of motion $u^{\mu}\nabla_{\mu} u^{\nu}= a^{\nu}$ for uniformly accelerated observer with initial data $( x(0)=x_{0},\,\dot{x}(0)=0)$. The equation of trajectories for two uniformly accelerated detectors with this initial data is given by,
\begin{subequations}
    \begin{eqnarray}
    x_{1}(\tau_{1}) &=& x_{01} + \frac{1}{a} - \frac{1}{a} \cosh (a \tau_{1})\,,\\
    x_{2}(\tau_{2}) &=& x_{02} + \frac{1}{a} - \frac{1}{a} \cosh (a \tau_{2})~.
\end{eqnarray}
\end{subequations}
It should be noted that for both detectors we have considered the same acceleration but with different separations along the $x$-axis at $\tau=0$. Following a similar procedure for calculations as done in Sec. \ref{subsec:Gauss-negativity}, we shall calculate the Negativity for Gaussian switching. In this regard, in a manner similar to the procedure adopted in Sec. \ref{sec:det-trjkt} and using the above equations for trajectories we calculate the geodesic distance between two points on different trajectories as 
\begin{equation}\label{sigma:non_rindler_mink}
    \Sigma^{2} = \frac{1}{a^{2}}\l[\l\{a\,d + \cosh(a\tau_{1})-\cosh(a\tau_{2})\r\}^{2} - \l\{\sinh(a \tau_{1})- \sinh(a\tau_{2})\r\}^{2}\r]~,
\end{equation}
where $d=x_{01}-x_{02}$. We substitute the above expression in Eq \eqref{eq:gen-Feynman} to arrive at the Feynman propagator. After performing the integral \eqref{eq:Ie-general} and utilizing the expressions of the individual transition probabilities of the uniformly accelerated detectors, we arrived at the Negativity which is depicted in Fig. \ref{fig:non-rindler-mink}.   

%%%%%%%%%%%%%%%%%%%%%%%%%%%%%%%%%%%%%%%%%%%%%%%%%%%%%%%%%%%%%%%%%%%%%%%%%
%%%%%%%%%%%%%%%%%%%%%%%%%%%%%%%%%%%%%%%%%%%%%%%%%%%%%%%%%%%%%%%%%%%%%%%%%
%
In the plots of Fig. \ref{fig:non-rindler-mink} it is easy to infer that for a given initial distance between detectors, the Negativity approaches the static detector limit as acceleration is decreased (red curves) and the peak in Negativity disappears. Whereas for a given acceleration $a$, the Negativity increases with decreasing initial distance $d$ between the detectors. This enhancement can be attributed to causal communication between the detectors. The Negativity curves seem to resemble Case I and show no new features. The role played by two different accelerations $a_{1}$ and $a_{2}$ in Case I is played by common acceleration $a$ and the initial distance $d$ between the detectors. We also note that if the acceleration is increased even for a very small separation $d$, there can be some peak in the Negativity, see the red solid curve. We would also like to mention that the Negativity curve for very small acceleration and separation between the detectors resembles the plot of Negativity for geodesic detectors with very small curvature, see the right plot of Fig. $5$ of \cite{K:2023oon}.

\subsection{de Sitter background}
\begin{figure}[h!]
\centering
\includegraphics[width=12.0cm]{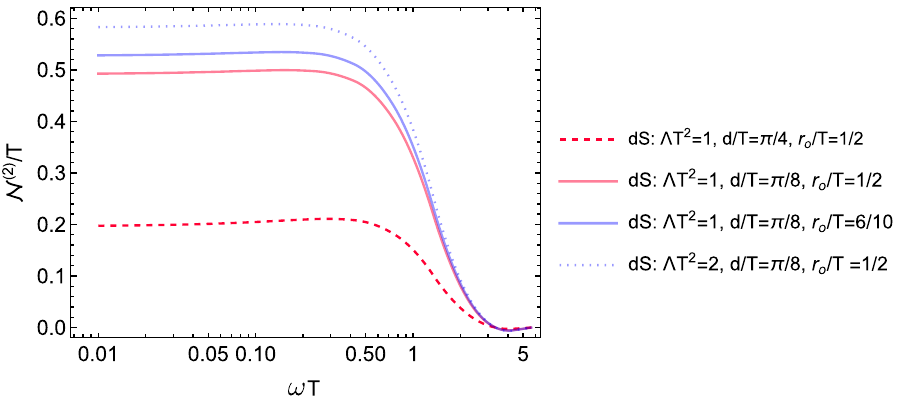}
\caption{The Negativity rate $(\mathcal{N}^{(2)}/T)$ is plotted against $\omega T$ between 2 trajectories with same acceleration $a$ separated by an initial distance $d$ at $\tau_{1}=\tau_{2}=0$ in de Sitter spacetime. Red curves show an increase in Negativity for a given de Sitter background ($\Lambda T^{2}=1$) with increasing initial separation ($d/T$). Blue curves show an increase in Negativity with increasing background curvature ($\Lambda T^{2}$) for a given initial separation $d$.}
\label{fig:non-rindler-ds}
\end{figure}

Second, we consider uniformly accelerated curves in de Sitter spacetime. The trajectories can be suitably written in terms of static coordinates $(t,r,\theta,\phi)$ with ($\theta_{0}=\pi/2$) as follows \cite{weinberg1972gravitation},
\begin{subequations}
\begin{eqnarray}
    T_{j} &=& \sqrt{\frac{1}{\Lambda} - r_{0}^{2}} \sinh(t_{j}\sqrt{\Lambda})\\
    X_{j} &=&  \sqrt{\frac{1}{\Lambda} - r_{0}^{2}} \cosh(t_{j}\sqrt{\Lambda})\\
    Y_{j} &=& r_{0}\cos(\theta_{0})\\
    Z_{j} &=& r_{0}\sin(\theta_{0})\cos\l(\phi_{0}+ \frac{d_{0}}{r_{0}}\,\delta_{j,2}\r)\\
    W_{j} &=& r_{0}\sin(\theta_{0})\sin\l(\phi_{0}+ \frac{d_{0}}{r_{0}}\,\delta_{j,2}\r)
\end{eqnarray}
\end{subequations}
where we have a rotation by an angle $d_{0}/r_{0}<\pi$ in the $(\theta,\phi)$ coordinates for the second trajectory with the same value of acceleration. It is straightforward to see that the corresponding geodesic distance is given by,
\begin{equation}\label{sigma:non-rindler-ds}
    \Sigma^{2} = \left( \frac{1}{\Lambda} - r_{0}^{2} \right) \left[ 2 - 2 \cosh\l\{\sqrt{\Lambda}(\tau_{2}-\tau_{1})\r\} + r_{0}^{2}\l\{2-2 \cosh(d_{0}/r_{0})\r\} \right]~.
\end{equation}
Since the above $\Sigma^{2}$ expression is only a function of $\tau_{1}-\tau_{2}$, we redefine integration variables as $u=\tau_{1}-\tau_{2}$ and $v=\tau_{1}+\tau_{2}$ which largely simplifies the calculation of $\mathcal{I}_{\epsilon}$ integral \eqref{eq:Ie-general} under Gaussian switching. And since individual detector transitions remain similar to previous cases, we calculated the Negativity and obtained the following results as presented in Fig. \ref{fig:non-rindler-ds}. 

From the above plots, we see an enhancement of Negativity with decreasing initial separation between the detectors for a given background curvature (red curves). At the same time, increasing background curvature enhances Negativity for a given initial separation between detectors (blue curves). Moreover, the Negativity curves agree with the geodesic detector limit as the curvature is decreased. We have seen that a bump appears in the Negativity when acceleration is increased in Minkowski spacetime.  

%As Einstein's equivalence principle requires, a similar bump should appear in Negativity in de Sitter spacetime for \emph{some} choice of $\Lambda$ and $d$, an absence of which implies a possible violation of equivalence principle. We have seen that background curvature does enhance Negativity, but it does not feature a peak. The difference in result can be understood by looking at \eqref{sigma:non_rindler_mink} and \eqref{sigma:non-rindler-ds} (under $r_{0}\longrightarrow 0$ limit). We see that $\Sigma^{2}_{dS} = \Sigma^{2}_{dS}(\Delta \tau)$, whereas $\Sigma^{2}_{M} = \Sigma^{2}_{M}(\tau_{1},\tau_{2})$, hence Feynman propagator have different pole structure implying different Negativity curves. We emphasize that the computation of entanglement is of a nonlocal nature in the sense that it requires two uniformly accelerated detectors in an extended region with the same acceleration, which requires a uniform gravitational field over an extended region and can never be mimicked by gravity. Whereas the equivalence principle only advocates for the replacement of uniform acceleration by free fall in the curved background over local regions only.

%%%%%%%%%%%%%%%%%%%%%%%%%%%%%%%%%%%%%%%%
%\bibliographystyle{apsrev}
%\bibliography{bibtex}

%%%%%%%%%%%%%%%%%%%%%%%%%%%%%%%%%%%%%%%%

\end{document}